\documentclass[review,10pt]{elsarticle}
\usepackage{lineno,natbib}

\usepackage{graphicx}
\usepackage{amsfonts}
\usepackage{amssymb}
\usepackage{float}
\usepackage{latexsym}
\usepackage{amsmath}
\usepackage{subfigure}
\usepackage[euler]{textgreek}
\usepackage{textcomp}
\usepackage{hyperref}
\usepackage{float}

\usepackage{algorithm}
\usepackage{algpseudocode}
\usepackage[table]{xcolor}

\newcommand{\pd}[2]{\frac{\partial #1}{\partial #2}}

\newcommand{\ub}[0]{\mathbf{u}}
\newcommand{\Db}[0]{\mathbf{D}}

\newcommand{\nb}[0]{\mathbf{n}}
\newcommand{\etal}{\textit{et al.~}}
\newcommand{\ie}{\textit{i.e.}}

\graphicspath{ {./HD_figures/} }

\usepackage{nomencl,etoolbox}

\makenomenclature

\newif\ifnomentry

\ExplSyntaxOn 
\NewExpandableDocumentCommand{\stringcase}{mm}
 {
  \str_case:nn { #1 } { #2 }
 }
\ExplSyntaxOff

\begin{document}
	\hypersetup{pdfauthor={Bradley Boyd}}

	\begin{frontmatter}
		
		\title{Simulation and modeling of the vaporization of a freely moving and deforming drop at low to moderate Weber numbers}
		
		\author[add1]{Bradley Boyd\corref{cor1}}
		\ead{bradley.boyd@canterbury.ac.nz}
		\author[add1]{Sid Becker}
		\ead{sid.becker@canterbury.ac.nz}
		\author[add2]{Yue Ling}
		\ead{stanley_ling@sc.edu}
		\cortext[cor1]{Corresponding author. 
			Address:    
			Private Bag 4800, Christchurch 8140, New Zealand
		} 
		\address[add1]{Department of Mechanical Engineering, University of Canterbury, Private Bag 4800, Christchurch 8140, New Zealand}
		\address[add2]{Department of Mechanical Engineering, University of South Carolina, 300 Main St, Columbia, SC 29208, USA}
		
		\begin{abstract}
The vaporization of a freely moving drop in a uniform, high-temperature gas stream is investigated through direct numerical simulation. The incompressible Navier-Stokes equations with surface tension and phase change are solved in conjunction with the energy equations of each phase. The sharp liquid-gas interface is tracked using the geometric Volume-of-Fluid (VOF) method and an immersed Dirichlet boundary condition for temperature is imposed at the interface. The simulation approach is validated by simulating water and acetone drops at nearly zero Weber numbers, and the simulation results agree very well with the empirical relation for spherical drops. Parametric simulations were conducted to investigate the aerodynamic breakup of vaporizing drops at low to moderate Weber and Reynolds numbers. The range of Weber numbers considered has covered the vibrational and bag breakup regimes. Through the simulation results, we have characterized the impact of drop deformation and breakup on the drop vaporization rate. When the drop Weber number increases, the windward surface area increases more rapidly over time. As a result, the rate of drop volume reduction also increases. The correlation between the vaporization rate and the windward surface area is examined for different Weber and Reynolds numbers. Using the approximate correlation between the drop vaporization rate and the windward surface area and the TAB model for drop deformation, a new time-dependent drop vaporization model is proposed. The present model agrees well with the simulation results and shows a significant improvement over the conventional model for spherical drops.
		\end{abstract}

		
	\end{frontmatter}



\nomenclature[C]{$a$}{Drop width along the axis of symmetry}
\nomenclature[C]{$D$}{Diameter}
\nomenclature[C]{$R_0$}{Spherical drop radius}
\nomenclature[C]{$R$}{Lateral drop radius}
\nomenclature[C]{$x,y,z$}{Cartesian coordinates}
\nomenclature[C]{$r,z$}{Axisymmetric cylindrical coordinates}
\nomenclature[C]{$\mathbf{g}$}{Gravity vector}
\nomenclature[C]{$\text{CFL}$}{Courant–Friedrichs–Lewy condition}
\nomenclature[C]{$\text{Oh}$}{Ohnesorge number}
\nomenclature[C]{$\text{Nu}$}{Nusselt number}
\nomenclature[C]{$\text{Nu}_s$}{Empirical Nusselt number for a spherical drop}
\nomenclature[C]{$\text{Re}$}{Reynolds number}
\nomenclature[C]{$\text{M}$}{Viscosity ratio}
\nomenclature[C]{$\text{Pr}$}{Prandtl number}
\nomenclature[C]{$\mathbf{u}$}{Velocity vector}
\nomenclature[C]{$p$}{Pressure}
\nomenclature[C]{$T$}{Temperature}
\nomenclature[C]{$k$}{Thermal conductivity}
\nomenclature[C]{$h_{l,g}$}{Specific latent heat of vaporization}
\nomenclature[C]{$C_p$}{Isobaric-specific heat}
\nomenclature[C]{$U$}{Velocity magnitude}
\nomenclature[C]{$t$}{Time}
\nomenclature[C]{$Z_f$}{Empirical correlation expression}
\nomenclature[C]{$s_{\gamma}$}{volumetric source due to mass transfer between phases}
\nomenclature[C]{$f$}{Liquid volume fraction}
\nomenclature[C]{$j_{\gamma}$}{Mass flux at the interface due to phase-change}
\nomenclature[C]{$V_l$}{Drop volume}
\nomenclature[C]{$\dot{V_l}$}{Drop vaporization rate}
\nomenclature[C]{$A$}{Area}
\nomenclature[C]{$C_b,C_d,C_f,C_k$}{Taylor Analogy Breakup parameters}
\nomenclature[c]{$L$}{Level of mesh refinement}
\nomenclature[c]{$l$}{Computational domain length}
\nomenclature[G]{$\eta$}{Density ratio}
\nomenclature[G]{$\mu$}{Viscosity}
\nomenclature[G]{$\tau_c$}{Characteristic breakup time}
\nomenclature[G]{$\rho$}{Density}
\nomenclature[G]{$\sigma$}{Surface tension coefficient}
\nomenclature[G]{$\kappa$}{Interface curvature}
\nomenclature[G]{$\mathbf{n}_{\gamma}$}{Interface normal}
\nomenclature[G]{$\delta_{\gamma}$}{Interface Dirac distribution function}
\nomenclature[G]{$\phi_{\gamma}$}{Interfacial area density}
\nomenclature[X]{$*$}{Non-dimensional variables}   
\nomenclature[Y]{$0$}{Initial value}
\nomenclature[Y]{$l$}{Liquid}
\nomenclature[Y]{$cr$}{Critical}
\nomenclature[Y]{$e$}{Ellipsoid}
\nomenclature[Y]{$g$}{Gas}
\nomenclature[Y]{$f$}{Final}
\nomenclature[Y]{$\text{sat}$}{Saturation}
\nomenclature[Y]{$\infty$}{Far-field}
\nomenclature[Y]{$\text{rel}$}{Relative}
\nomenclature[Y]{$\gamma$}{Interfacial}
\nomenclature[Y]{$s$}{Spherical drop}
\nomenclature[Y]{$max$}{Maximum}
\nomenclature[Y]{$w$}{Windward}

\printnomenclature

\section{Introduction}

Vaporization of freely moving drops is commonly seen in a wide variety of industrial applications such as liquid fuel injection and spray cooling. However, the interaction between drop aerobreakup and vaporization is not fully understood.  To investigate the fundamental multiphase flow dynamics and the phase change involved, drop aerobreakup is often studied in an idealized configuration, where a stationary drop is suddenly subjected to a uniform gaseous stream at a constant velocity \cite{theofanous_aerobreakup_2011, marcotte_density_2019}. The morphological change of the drop is controlled by the competition between the destabilizing inertia force of the gas stream and the stabilizing forces, including the surface tension on the gas-liquid interfaces and the liquid viscous forces. Therefore, the aerodynamic deformation and breakup of drops are fully determined by four independent dimensionless parameters: the Weber ($\text{We}_0$), Reynolds ($\text{Re}_0$), and Ohnesorge ($\text{Oh}$) numbers, and the liquid-to-gas density ($\eta$) ratio \cite{guildenbecher_secondary_2009, jain_secondary_2019, marcotte_density_2019}, which are defined as
\begin{align}
&\text{We}_0=\frac{\rho_g U_\infty^2 D_0}{\sigma},\, \; 
\text{Re}_0=\frac{\rho_g U_{\infty} D_0}{\mu_g}, \; 
\text{Oh} = \frac{\mu_l}{\sqrt{\rho_l \sigma D_0 }}, \;
\eta=\frac{\rho_l}{\rho_g}, \;
\label{eq:non-dim_numbers}
\end{align}
where $\mu$, $\rho$, $\sigma$, $D_0$, and $U_\infty$ are the dynamic viscosity, density, surface tension coefficient, initial drop diameter, and the far-field velocity, respectively. Note that the subscripts $l$ and $g$ denote the properties of the liquid and gas. Alternative dimensionless parameters can be defined based on the above four parameters, such as the liquid-to-gas viscosity ratio $\textsl{M}=\mu_l/\mu_g$ \citep{guildenbecher_secondary_2009}.

Most former studies on drop aerobreakup are through experiments \cite{ranger_aerodynamic_1969, gelfand_singularities_1975, hsiang_near-limit_1992, theofanous_physics_2008, sharma_shock_2021}, though numerical studies using high-fidelity interface-resolved simulations are emerging in recent years \cite{meng_numerical_2018, marcotte_density_2019, jain_secondary_2019}. A variety of breakup modes, including the vibrational, bag, multi-mode, and shear modes, have been observed in both experiments and simulations when the Weber number is varied. The critical Weber number $(\text{We}_{cr})$ refers to the minimum Weber number at which breakup first occurs. In the present study, we consider ``low Weber numbers" as $\text{We} < \text{We}_{cr}$, while ``moderate Weber numbers" are for $\text{We}_{cr}< \text{We} \lesssim 120$, for which the drop breaks in the bag or multi-mode regimes \cite{jain_secondary_2015}. The present study does not consider cases with very high $\text{We}$ that are in the shear regime, due to the extreme mesh resolution required to fully resolve the interfacial dynamics in that regime \cite{chang_direct_2013, meng_numerical_2018}.

When the drop is exposed to a high-temperature gas stream, heat and mass transfer occur during the aerobreakup process, and additional parameters are introduced. In the present study, the vaporization of a spherical drop is driven by the heat transfer between the hot gas stream and the drop, which is thus characterized by the Nusselt number ($\text{Nu}$)
\begin{align}
	\text{Nu} = \frac{h D}{k_g}, \; h=\frac{q}{(T_\infty - T_\text{sat})}
\end{align}
where $h$, $q$, $T_{\text{sat}}$, $T_\infty$ is the heat transfer coefficient, heat flux, saturation temperature, and far-field temperature, respectively. The Nusselt number depends on $\text{Re}$ as well as the Stefan and Prandtl numbers, which are defined as
\begin{align}
\text{St} = \frac{C_{p,g}(T_{\infty}-T_\text{sat})}{h_{l,g}}, \;
\text{Pr} =\frac{C_{p,g} \mu_{g}}{k_g}.
\end{align}
where $k$, $C_p$, and $h_{l,g}$ are the thermal conductivity, the isobaric-specific heat, and the specific latent heat of vaporization, respectively.

Experimental studies have extensively investigated the vaporization of a spherical drop in high-temperature environments \cite{renksizbulut_experimental_1983, sazhin_advanced_2006, haywood_detailed_1989, chiang_numerical_1992, yuen_heat-transfer_1978}. To maintain the drop surface's sphericity, a liquid film was applied to a porous sphere instead of an actual drop, with liquid continuously supplied to the sphere. The perfectly spherical interface of the liquid drop at saturation temperature is then exposed to a high-temperature stream, resulting in vaporization. The rate of liquid vaporization is therefore equal to the rate of liquid supplied to the sphere, keeping the radius of the spherical surface unchanged. Empirical data were typically condensed into empirical relations of the Nusselt number expressed as a function of the Reynolds, Prandtl, and Stefan numbers, such as the commonly-used Renksizbulut-Yuen (RY) model \citep{renksizbulut_experimental_1983}. 

The RY correlation is only strictly valid for spherical drops, namely in the limit of zero Weber number. For drops with finite Weber numbers, the vaporization rate is influenced by the deformation of the drop. In the last decade, multiple numerical studies have been reported in the literature on the vaporization of deformable fuel drops, where the drops are falling \cite{reutzsch_consistent_2020, irfan_front_2017} or impulsively accelerated \cite{jin_numerical_2019, banerjee_numerical_2013, strotos_numerical_2011, zhao_numerical_2014, jin_numerical_2018, raghuram_numerical_2012}. However, these previous studies focused on low Weber numbers, where the drops only experienced mild deformation. Recently, Setiya \etal \cite{setiya_quasi-steady_2023} investigated the evaporation of deforming drops at low Weber and Reynolds numbers and observed an increase in the evaporation rate with drop deformations.

Numerical studies of vaporization of drops at moderate Weber numbers, for which the drops undergo significant deformation and even breakup, did not emerge until very recent numerical studies \cite{gao_effect_2022, boyd_consistent_2023}: this is because sophisticated numerical methods are required to rigorously resolve the sharp and vaporizing interface, and typically a high mesh-resolution is needed to well resolve the thermal and velocity gas boundary layer near the drop surface to accurately predict the drop vaporization rate. In the recent work by Boyd and Ling \cite{boyd_consistent_2023}, a novel numerical method for implementing phase change with the volume-of-fluid (VOF) method was proposed. This method enables direct numerical simulation of interfacial multiphase flows with phase change. The volumetric source due to vaporization was added to the pressure equation to account for the non-zero divergence of velocity at the interface and the resulting Stefan flow. To prevent perturbations of the Stefan flow on the velocity at interfacial cells, a new treatment was proposed to handle vaporization-induced volumetric sources. This treatment ensures that the velocities at the interface cells are accurately represented and can be directly used in VOF advection.

Accurate simulation of the vaporization of deformable drops is important for the development of Lagrangian point-particle (LPP) models for drops, which are typically required for large-scale numerical simulations of sprays. Since it is too expensive to resolve the flows and interfacial dynamics for each individual drop, the drops are treated as point particles, and the drop vaporization rate is estimated by physical models \cite{sirignano_fuel_1983}, similar to those for drag and heat transfer \cite{balachandar_turbulent_2010, ling_scaling_2013, ling_inter-phase_2016}. As explained above, when the vaporization is driven by the heat transfer between the drop and the hot gas stream, the drop vaporization rate is represented by the Nusselt number. Conventionally, the drop vaporization is modeled as a quasi-steady process, such as the RY correlation, where the Nusselt number depends on the instantaneous Reynolds, Prandtl, and Stefan numbers, which are calculated based on the undisturbed gas flow properties, the drop velocity, and temperature at the current time. However, for finite Weber numbers, the drop shape changes over time, which in turn introduces a transient effect on the drop vaporization rate. Therefore, similar to the models for drop deformation, such as the TAB model \cite{orourke_tab_1987}, which predicts the drop shape evolution over time, the vaporization model also needs to incorporate the transient effect. To the best of the authors' knowledge, such a transient vaporization model for deformable drops with finite Weber numbers has not been developed in the literature.

The objectives of the present study are to characterize the evolution of the vaporization rate through numerical simulations and to extend the drop vaporization model by incorporating the time-dependent drop deformation. The present study will focus on drops at low and moderate Weber and Reynolds numbers, for which 2D axisymmetric simulations are sufficient to resolve the drop deformation until the breakup occurs, and the relatively low computational costs will allow parametric simulations of a large number of cases. The extension of the present simulations to high Weber and Reynolds numbers will require high-resolution 3D simulations, which will be more computationally expensive and will be the subject of future work. Since the numerical methods used for the present simulations have been introduced in a previous paper \cite{boyd_consistent_2023}, only a brief summary will be given here (see Section \ref{section:numerical_model}). In the results, we will first validate the solver by simulating a spherical drop in the limit of vanishing Weber number and compare the results against the empirical correlation (Section \ref{section:validation}). The results for low and moderate Weber number cases are then presented in Sections \ref{section:low_We} and \ref{section:moderate_We}. The relation between drop deformation, characterized by the windward surface area, and the drop vaporization rate will be investigated. The effect of the Reynolds number will be examined in Section \ref{section:influence_of_Re}. Full 3D simulations for several representative cases are performed, and the results are compared with those obtained by 2D axisymmetric simulations to confirm that 2D simulation is sufficient to capture the correlation (see Section \ref{section:3D_section}). Finally, we present the new time-dependent drop vaporization model based on the correlation between drop deformation and vaporization and the TAB drop deformation model in Section \ref{section:VTAB}.

\section{Simulation methods}
\label{section:numerical_model}
\subsection{Governing equations}
The incompressible two-phase interfacial flows with vaporization are governed by the 
Navier-Stokes equation with surface tension, 
\begin{align}
	& \rho \left( \frac{\partial \ub}{\partial t}+ \ub \cdot \nabla \ub \right)= - \nabla p + \nabla \cdot (2\mu \Db) + \rho \mathbf{g}+ \sigma \kappa \delta_{\gamma} \nb_{\gamma}
	\label{eq:momentum}
\end{align}
where $\mathbf{u}$, $p$, $\mu$, $\rho$, $\sigma$, $\kappa$, $\mathbf{n}_{\gamma}$, and $\delta_{\gamma}$ are the velocity, pressure, dynamic viscosity, density, surface tension coefficient, curvature, interface normal, and interface Dirac distribution function, respectively.  The deformation tensor is defined as $\Db = (\nabla \ub + \nabla \ub^T)/2$. Gravity, which can be easily included, is neglected in the present study. 

The one-fluid approach is adopted, in which the two phases are treated as one fluid with properties changing abruptly across the interface. The two different phases are distinguished by the liquid volume fraction, $f$, so $f=1$ and 0 for cells with purely liquid and gas, respectively, and $0<f<1$ for cells containing the interface and both gas and liquid. The evolution of $f$ follows the advection equation,
\begin{align}
	\frac{\partial f}{\partial t}+ \nabla \cdot \left( f \mathbf{u}\right ) = \frac{-s_{\gamma}}{\rho_l}
	\label{eq:vof_advection}
\end{align}
where the subscripts $l$ and $g$ denote the liquid and gas properties. The source term on the right accounts for the additional change in the interface location due to phase change, and $s_{\gamma}$ is the volumetric source due to phase change, which depends on the mass flux at the interface ($j_{\gamma}$) and the interfacial area density ($\phi_{\gamma}$) as
\begin{align}
	s_{\gamma} = j_{\gamma} \phi_{\gamma}
	\label{eq:sm}
\end{align}
 The interfacial area density is $\phi_{\gamma}=A_{\gamma}/V_c$, where $A_{\gamma}$ is the liquid-gas interface area in a finite-volume cell with a volume $V_c$.  The mass flux due to phase change, 
\begin{align}
	j_{\gamma} =\frac{1}{h_{l,g}}\left(k_l \nabla T|_{l,\gamma} \cdot \mathbf{n}_{\gamma} - k_g \nabla T|_{g,\gamma} \cdot \mathbf{n}_{\gamma}\right)\,,
	\label{eq:j_gamma}
\end{align}
is driven by heat transfer, where $T$, $k$, and $h_{l,g}$ are the temperature, thermal conductivity, and specific latent heat of vaporization, respectively. The gas and liquid temperature fields are required to calculate $j_\gamma$, which are in turn obtained by solving the energy conservation equation for both the liquid and gas phases \cite{boyd_consistent_2023,sato_sharp-interface_2013, bures_direct_2021, malan_geometric_2021}
\begin{align}
	& \rho_g C_{p,g} \left( \pd{T_g}{t} +  \ub \cdot \nabla T_g  \right) = \nabla \cdot (k_g \nabla T_g) \, ,
	\label{eq:temp_gas}\,\\ 
	& \rho_l C_{p,l}\left(\pd{T_l}{t} +   \ub \cdot \nabla T_l \right) = \nabla \cdot (k_l \nabla T_l) \, 
	\label{eq:temp_liq}
\end{align}
where $C_p$ is the isobaric-specific heat. Since $T_g$ and $T_l$ are only solved in the gas and liquid regions, the gas-liquid interface is treated as an embedded boundary where the temperature remains as the saturation temperature, $T_{\text{sat}}$.

Finally, the phase change also results in a modification of the continuity equation, 
 \begin{align}
    \nabla \cdot \mathbf{u}= s_{\gamma} \left( \frac{1}{\rho_g}-\frac{1}{\rho_l} \right)\,.
    \label{eq:divergence}
\end{align}

\subsection{Numerical methods and solvers} 
The governing equations are solved using a finite volume approach. The advection of the liquid volume fraction (Eq.~\eqref{eq:vof_advection}) is solved using a geometric VOF method \cite{weymouth_conservative_2010}. The projection method incorporates the incompressibility condition in the momentum equation, and the pressure Poisson equation is solved using the multi-grid method. The advection of momentum near the interface is conducted in a manner consistent with the VOF advection \cite{boyd_consistent_2023}. The surface tension term in the momentum equation is discretized using the balanced-force continuum-surface-force method \cite{francois_balanced-force_2006}. The interfacial curvature that is required for surface tension is calculated using the height-function method \cite{popinet_accurate_2009}. A second-order staggered-in-time discretization of the volume fraction and pressure is used. The quadtree/octree mesh discretizes the 2D/3D spatial domains, providing important flexibility to dynamically refine the mesh in user-defined regions. The adaptation criterion is based on the wavelet estimate of the discretization errors of the liquid volume fraction, temperature, and all velocity components. 

The energy equations for both phases (Eqs.~\eqref{eq:temp_gas}, \eqref{eq:temp_liq}) are solved with the Dirichlet boundary condition at the interface, \ie, $(T_g)_\gamma=(T_l)_\gamma = T_{\text{sat}}$. The advection terms are treated similarly to momentum and are consistent with VOF advection, while the diffusion terms are implicitly integrated using the multi-grid approach developed by \cite{lopez-herrera_electrokinetic_2015} and subsequently used for diffusion-driven evaporation problems in \cite{gennari_phase-change_2022}. 

\begin{figure}[tbp]
	\begin{center}
		\includegraphics [width=1.0\columnwidth]{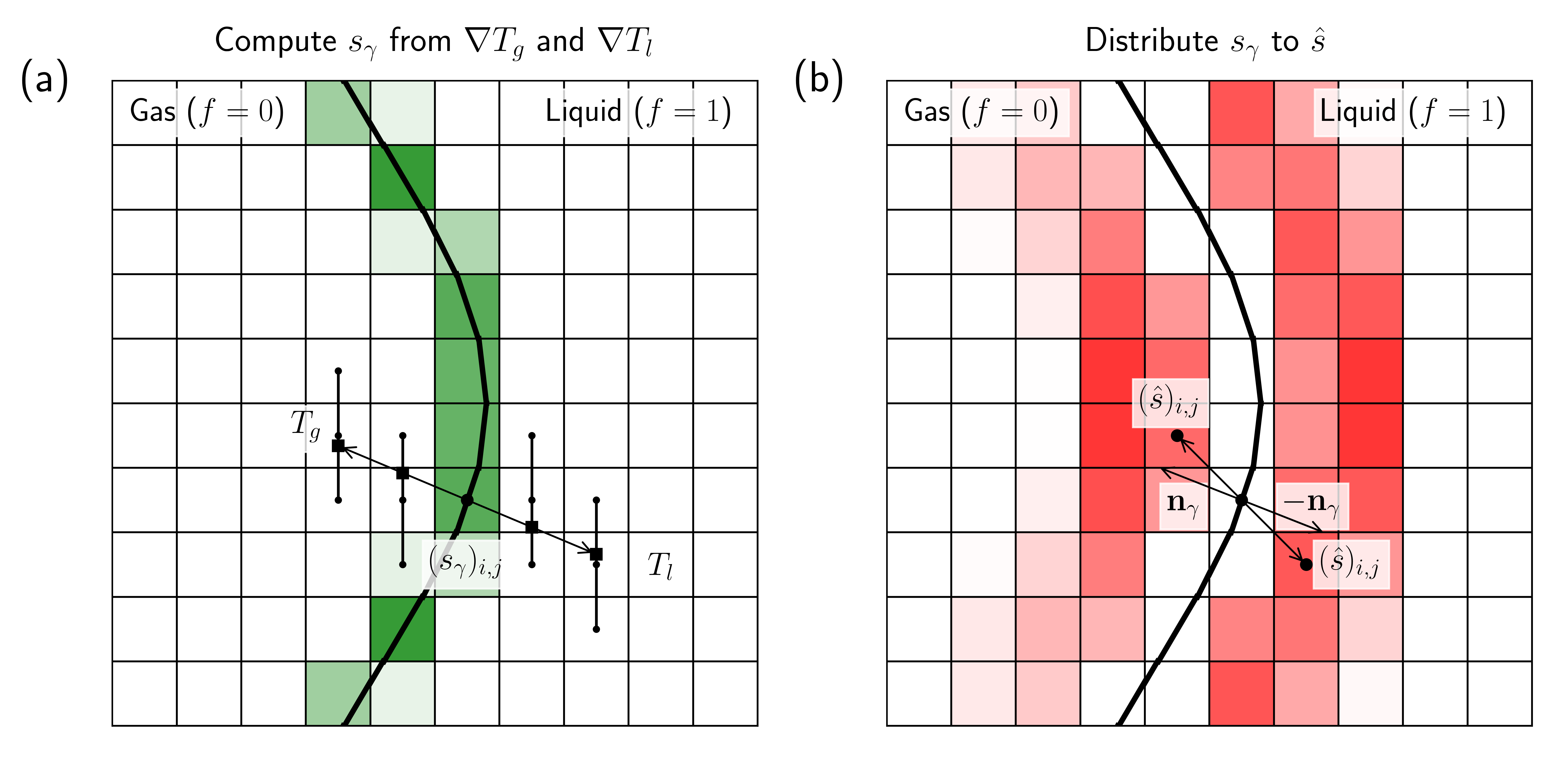}
	\end{center}
	\caption{A schematic of the numerical methods for (a) determining the volumetric vaporization source term ($s_\gamma$) at the interface and (b) distributing $s_\gamma$ to the near pure liquid and gas computational cells ($\hat{s}$) to ensure a divergence-free flow at the interfacial cells for the VOF advection procedure.}
	\label{fig:numerical_methods} 
\end{figure}

The determination of $j_\gamma$ (Eq.~\eqref{eq:j_gamma}) requires the temperature gradient on both the liquid and gas sides of the interface, \ie, $(\nabla T_g)_{\gamma}$ and $(\nabla T_l)_{\gamma}$. To avoid calculating the gradient across the interface, the temperature gradient for each phase is estimated using a three-point stencil in the normal direction from the reconstructed VOF interface \cite{johansen_cartesian_1998, schwartz_cartesian_2006, gennari_phase-change_2022}, see Fig. \ref{fig:numerical_methods}(a).

As the projection method is used to incorporate the continuity equation, the pressure Poisson equation is solved,
\begin{align}
    \nabla \cdot \left( \frac{\Delta t}{\rho} \nabla p \right) = \nabla \cdot \ub^{*} - \hat{s} \left( \frac{1-f}{\rho_g}-\frac{f}{\rho_l} \right) \,,
    \label{eq:poisson0}
\end{align}
where $\ub^{*}$ is the auxiliary velocity that accounts for all the terms in the momentum equation except the pressure term. The second term on the right is the additional source term induced by phase change, and $\hat{s}$ is equal to the sum of the source distribution from all the nearby interfacial cells. Instead of applying the volumetric source $s_\gamma$ directly at the interfacial cell, we distribute the volumetric source to the nearest pure gas cells in a $5^3$ stencil in 3D ($5^2$ in 2D), as described in \cite{boyd_consistent_2023}. Furthermore, we distribute $s_\gamma$ to the nearest pure liquid cells using a similar procedure (see Fig \ref{fig:numerical_methods}(b)). As a result, the integration of the distributed source $\hat{s}$ in all pure liquid cells $(f=1)$ and all pure gas cells $(f=0)$ are both equal to the integration of the volumetric source $s_\gamma$ over all the interfacial cells, \ie, $\int s_\gamma \,dV  = \int \hat{s}f \,dV = \int \hat{s}(1-f) \,dV$. By distributing the volumetric source from the interfacial cell to the neighboring pure gas and liquid cells, the velocity in the interfacial cell will not be ``contaminated" by the Stefan flow and will remain as the liquid velocity by which the interface moves. Previous studies have used a similar approach \cite{hardt_evaporation_2008, gao_effect_2022} but using a diffusion procedure.

The above numerical methods have been implemented in the open-source solver \emph{Basilisk} \cite{popinet_gerris_2003, popinet_accurate_2009, boyd_consistent_2023}. Validation studies of the present methods on two-phase interfacial flows with phase change can be found in our previous studies. Further validation specifically for drop vaporization is conducted in the present study and the results will be presented in Section~\ref{section:validation}. 

\section{Problem description}
A freely moving liquid drop with diameter $D_0$ is initially stationary and at saturation temperature in an unbounded domain of vapor. The drop is suddenly exposed to a uniform superheated stream of vapor with temperature $T_{\infty}$ and velocity $U_{\infty}$ at $t=0$. Note that the drop liquid and surrounding vapor are the same species; \ie, acetone liquid and vapor. The $x$ coordinate is chosen to be aligned with the free stream. Under the non-zero relative velocity between the drop and the free stream, \ie, $U_{\text{rel}} = U_\infty-u_d$, where $u_d$ is the mean velocity of the drop, the drop will be accelerated along the streamwise direction while it deforms and vaporizes. Therefore, the problem is an initial-value problem where the instantaneous relative velocity $U_{\text{rel}}$ and drop diameter $D$ both vary over time.  

As discussed in the introduction, the aerodynamic deformation and breakup of a drop are fully characterized by the key dimensionless parameters, the Weber ($\text{We}_0$), Reynolds ($\text{Re}_0$), and Ohnesorge ($\text{Oh}$) numbers, defined based on the initial drop diameter $D_0$ and the far-field velocity $U_\infty$ (which is also the initial relative velocity), as well as the liquid-to-gas density ratio ($\eta$). While the gas dynamic pressure tends to destabilize the drop, the surface tension is an important force to resist the deformation. The ratio between the two is represented by the Weber number, which is most commonly used to characterize drop aerobreakup regimes \cite{gelfand_singularities_1975, hsiang_near-limit_1992, theofanous_physics_2008}. The liquid viscosity is an additional resistant force \cite{theofanous_physics_2012}, and the relative importance between the liquid viscous force and surface tension is indicated by the Ohnesorge number. For the present study, we only consider drops with low $\text{Oh}$, so the surface tension is the dominant force that hinders drop deformation and breakup. The viscosity ratio is an alternative for $\text{Oh}$ to characterize the effect of liquid viscosity. The density ratio is also important to the early-time acceleration and deformation of the drop when the gas stream is in a high-pressure environment and $\eta$ is moderate \cite{marcotte_density_2019, jain_secondary_2019}. In the present study, we mainly focused on drops with high $\eta$. The Reynolds number characterizes the gas flow around the drop, in particular the wake dynamics, which in turn influences the drop deformation and vaporization. While $\text{We}_0$ and $\text{Re}_0$ with the subscript $0$ are defined based on the initial relative velocity and drop diameter (Eq. \eqref{eq:non-dim_numbers}). One can also define the counterparts based on the instantaneous relative velocity and the volume-based diameter $D=(6V_l/\pi)^{1/3}$, where $V_l$ is the instantaneous drop volume, \ie,  
\begin{align}
\text{We}=\frac{\rho_g U_\text{rel}^2 D}{\sigma}, 
\text{Re}=\frac{\rho_g U_\text{rel} D}{\mu_g}. 
\end{align} 

The dimensionless variables are defined based on the initial drop diameter and free stream conditions. The dimensionless position, velocity, and drop volume are defined as 
\begin{align}
x^* = \frac{x}{D_0}, \quad u^* = \frac{u}{U_\infty}, \quad V_l^*=\frac{V_l}{\pi D_0^3/6}. 
\label{eq:V_star}
\end{align}
Dimensionless time ($t^*$), which is based on the characteristic breakup time ($\tau_c$), is commonly used in the literature \cite{ranger_aerodynamic_1969}:
\begin{align}
t^* = \frac{t}{\tau_c}, \; \tau_c=\frac{D_0\sqrt{\eta}}{U_{\infty}}.
\label{eq:t_star}
\end{align}

\subsection{Simulation setup}
\begin{figure}[tbp]
	\begin{center}
		\includegraphics [width=0.7\columnwidth]{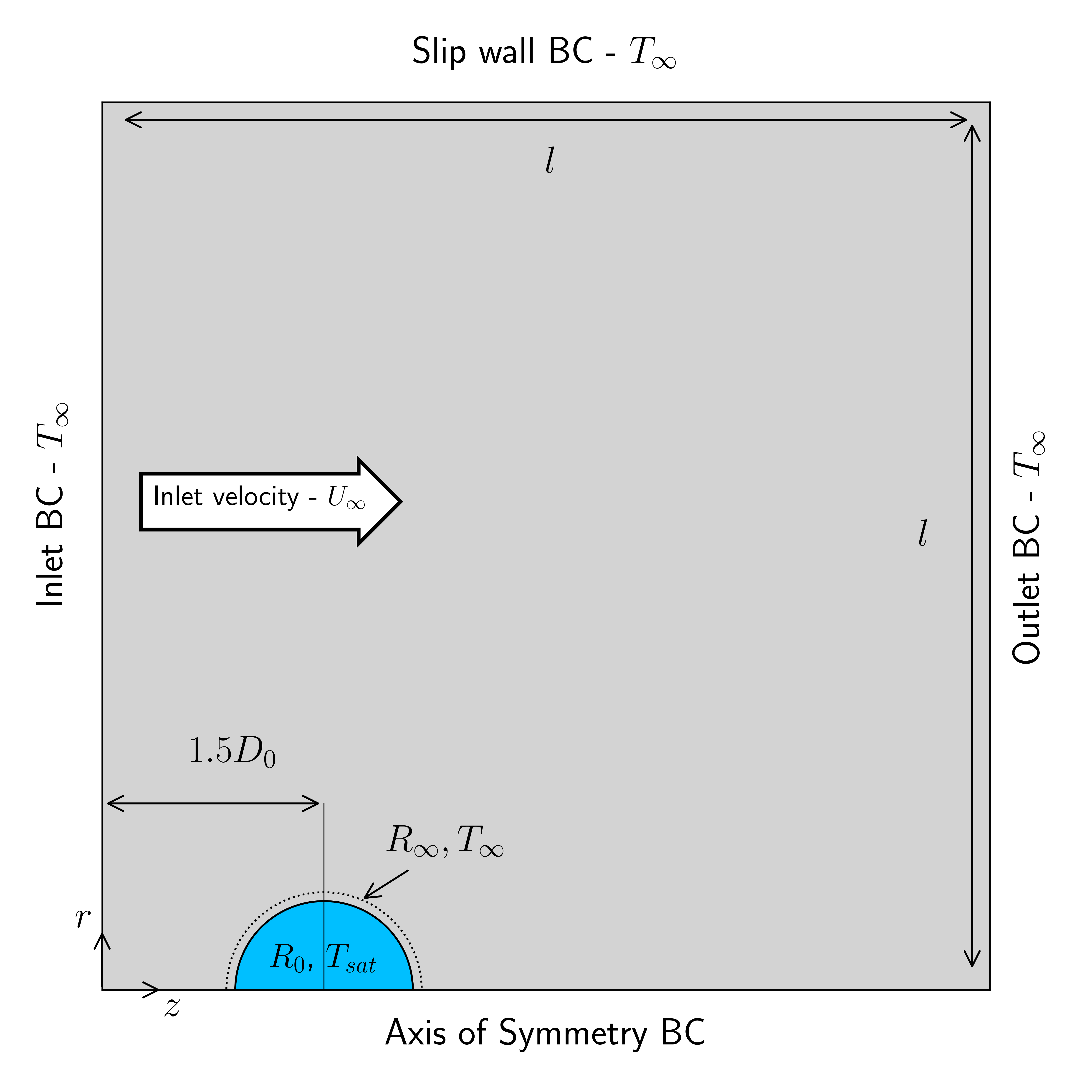}
	\end{center}
	\caption{The computational domain for 2D-axisymmetric simulation of the vaporization of a freely moving drop in a hot vapor stream. }
	\label{fig:moving_drop_diagram} 
\end{figure}

Two-dimensional axisymmetric simulations are performed for the present problem. The axisymmetric assumption is verified through full 3D simulations. The agreement between the results of 3D and 2D axisymmetric simulations is remarkably good within the range of $\text{Re}$ considered, which will be shown later in section \ref{section:3D_section}. Numerical and experimental investigations of falling drops conducted by Zhang et al. \cite{zhang_short-term_2019} indicated that 2D axisymmetric simulation results for drop deformation and velocity align well with experimental measurements up to $\text{Re}=600$. Although axisymmetry of the wake tends to break down for $\text{Re}\gtrsim 300$, it is important to note that the 3D wake might not be fully developed and could remain approximately 2D axisymmetric during the time scale in which the drop undergoes aerodynamic breakup. The computational domain is shown in Fig.~\ref{fig:moving_drop_diagram}. The domain is a square with an edge length of $l$. The drop is initially located at $x_0=1.5D_0$. Sensitivity analysis has been performed for different initial positions of the drop, confirming that the value of $x_0$ used is sufficiently large to eliminate the effect of the inlet boundary condition. The adaptive quadtree mesh was used to discretize the 2D domain. The mesh resolution is controlled by the maximum level of refinement $L$, which corresponds to $\Delta x = l/(2^L)$. The dynamic mesh adaptation is based on a wavelet-estimated discretization error \cite{popinet_quadtree-adaptive_2015, van_hooft_towards_2018}, where the refinement criteria are based on temperature, liquid volume fraction, and all velocity components. The time step is computed based on the CFL condition, and $\text{CFL} = 0.1$ for all cases. Two different liquids; water and acetone, are considered. The fluid properties are provided in Table~\ref{tab:properties}.

\begin{table*}[tb]
 \centering
 \begin{tabular}{l l l l l} 
     \hline
    Property    & \multicolumn{2}{c}{Water} & \multicolumn{2}{c}{Acetone} \\
      &     Liquid & Vapor &     Liquid & Vapor \\
     \hline
 $\rho$ $[kg/m^3]$ & $958.4$ & $0.597$ & $710$ & $5.11$  \\
 $k$ $[W \, m^{-1} \, K^{-1}]$ & $0.679$  & $0.025$ & $0.156$  & $0.0166$\\
 $C_{p}$  $[J \, kg^{-1} \, K^{-1}]$  & $4216$ & $2030$ & $2420$ & $1460$\\
  $\mu$ $[Pa \, s]$  & $2.8\text{e-}{4}$ & $1.26\text{e-}{5}$ & $1.85\text{e-}{4}$ & $9.59\text{e-}{6}$\\
 $h_{l,g}$  $[J \, kg^{-1}]$ & $2.26\text{e}{6}$ &- & $4.88\text{e}5$ &- \\
 $T_\text{sat}$  $[K]$ & $373.15$ &- & $359$ &- \\
 $\sigma$ $[N \, m^{-1}]$ & $0.0728$ &- & $0.0153$ &- \\
  $\text{Pr}$ & $1.023$ &- & $0.8435$ &- \\
     \hline
 \end{tabular}
 \caption{Properties of the saturated water and acetone, where water is used in Section \ref{section:validation} and acetone is used in Sections \ref{section:validation}-\ref{section:3D_section}.}
 \label{tab:properties}
 \end{table*}

The focus of the present study is to investigate the influence of drop deformation on the vaporization rate: the gas phase is considered to consist only of the vapor of the drop fluid. Indeed, this is a simplified model compared to the more general situation, where a drop vaporizes to a different gas. As the main contributions of the presence of another ambient gas to the drop deformation and vaporization lie at the gas mixture properties, like density and viscosity, the present setup can be viewed as the special case for which the ambient gas exhibits the same properties as the drop vapor. When the properties of the ambient gas and the drop vapor are different, the drop vapor concentration must be solved to accurately calculate the properties of the mixture. The variation of the gas properties will only influence the key dimensionless parameters, such as Weber, Reynolds, and Stefan numbers. For example, if an acetone drop is exposed to a stream of air instead of its vapor, since the air viscosity is higher than the acetone viscosity, and air density is lower than the acetone density, the Reynolds number for the same drop size and free stream velocity will be lower than the counterpart when the acetone drop is in its own vapor. Nevertheless, as the drop deformation/breakup and vaporization are dictated by the key dimensionless parameters, as long as the dimensionless parameters are the same, the present results remain valid in representing the physics of drops vaporizing into a different gas. The correlation between drop deformation and vaporization rate, as will be shown later, still applies. Detailed investigation of the mixing between drop vapor and ambient gas and its effect on drop breakup and vaporization will be a topic of our future work.

 
\section{Results}
\label{section:results}
\subsection{Validation studies: vaporization of approximately spherical drops}
\label{section:validation}
To validate the present simulation approach, drop vaporization for $\text{We}_0=0.5$ is studied. A parametric study has been performed by varying the Reynolds number, \ie, $22\leq \text{Re}_0 \leq 200$, and $0.1 \leq \text{St} \leq 2.05$. For each case, $U_\infty$ and $\sigma$ were modified to keep $\text{We}_0$ constant but vary $\text{Re}_0$. Due to the low $\text{We}_0$ and $\text{Re}_0$, drop deformation is very mild, and the drop shape is approximately spherical. As $\sigma$ changes, $\text{Oh}$ varies from about 0.01 to 0.2, but its effect on vaporization is negligible. The Stefan number, $\text{St}$, is changed by varying $T_{\infty}$.

For all cases in this section, the domain size is $l=8D_0$, and the mesh refinement level is $L=12$ (equivalent to $\Delta x = D_0/512$). The simulations are run until $t=6.4D_0/U_{\infty}$, which corresponds to $t^*\approx0.54$ for acetone and $t^*\approx0.17$ for water. After the velocity and thermal boundary layers fully develop, the flow and drop vaporization reach an approximate quasi-steady state. Until then, we measure the quasi-steady drop vaporization rate. To shorten the transition to the quasi-steady state, an initial temperature profile in the gas phase is specified around the drop surface in the thermal boundary layer as follows:
 \begin{align}
    T(r)= 
\begin{cases}
    T_\text{sat}\,,& \text{if } r\leq R_0, \\
    T_{\infty}\,,& \text{if } r\geq R_{\infty}, \\
    \frac{R-R_0}{R_{\infty}}(T_{\infty}-T_\text{sat})+ T_\text{sat}\,,& \text{otherwise.}
\end{cases}
 	\label{eq:init_temp_drop}
\end{align}
Here, $R_{\infty}=1.2R_0$ (see Fig.~\ref{fig:moving_drop_diagram}) and $R_0=D_0/2$ is the drop radius. Invoking this initial condition simply shortens the transition to the quasi-steady state. The specific value of $R_{\infty}$ is immaterial to the vaporization rate at the quasi-steady state, as confirmed by a sensitivity study on $R_{\infty}$ (see \ref{app:R_inf}). For all the cases in this section, a quasi-steady state was reached around $t^*=0.06$ and $t^*=0.4$ for the water and acetone drops, respectively.

\begin{figure}[tbp]
	\begin{center}
		\includegraphics [width=.8\columnwidth]{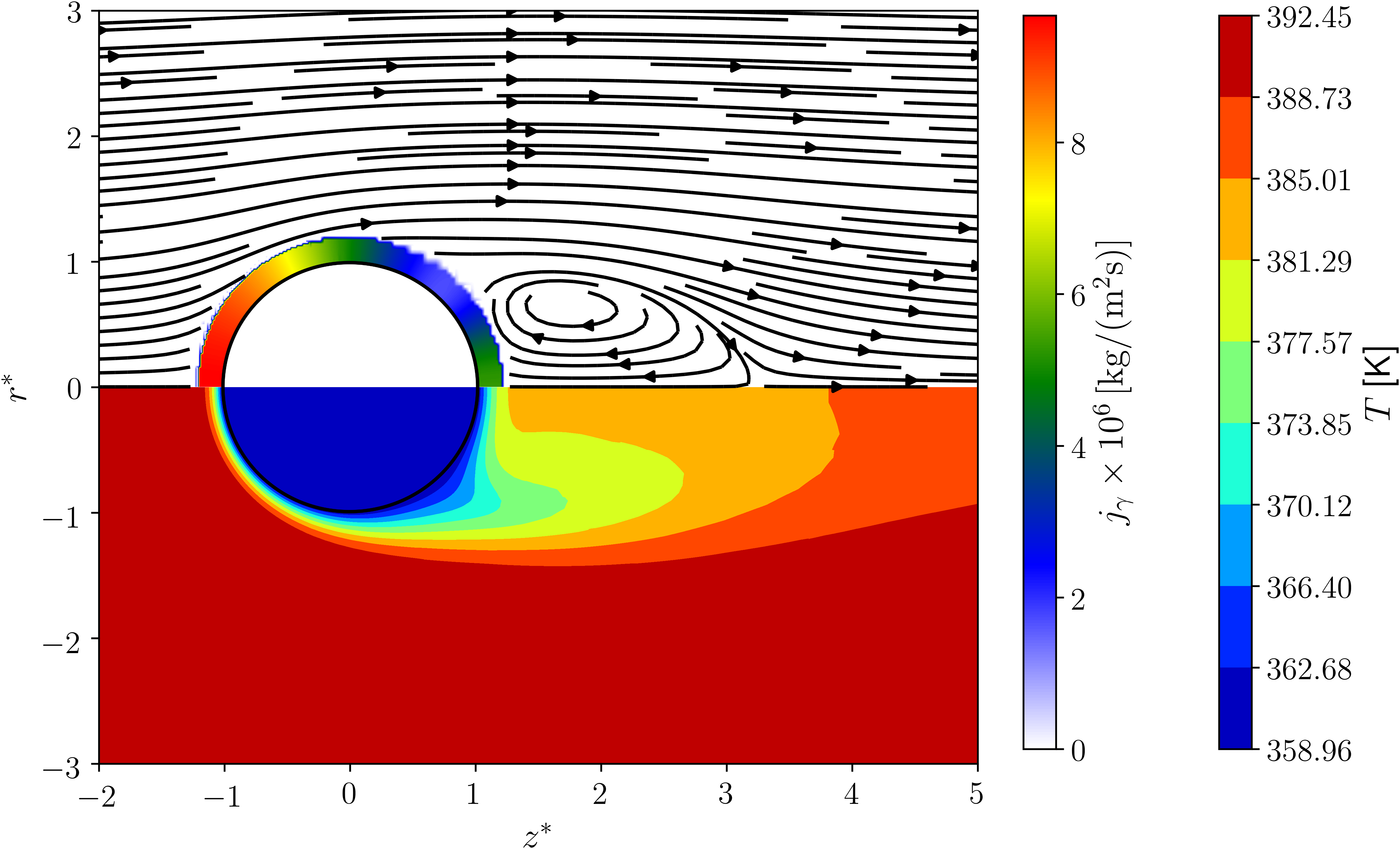}
	\end{center}
    \caption{A representative snapshot of the contour of the vaporization rate ($j_{\gamma}$), the flow streamlines (top), and the temperature field (bottom) for the acetone drop at $t^*=0.543$ for $\text{We}=0.5$, $\text{Re}=60$, and $\text{St}=0.1$. The dimensionless coordinates are $r^*=r/R_0$ and $z^*=z/R_0$. The vaporization region in the top figure has been ``thickened" for better visualization.}
	\label{fig:moving_drop_temperature} 
\end{figure}

Figure~\ref{fig:moving_drop_temperature} shows the temperature and velocity fields at the end of the simulation ($t^*=0.54$) for the acetone drop at $\text{Re}_0= 60$. A contour plot of $j_{\gamma}$ is used to show the vaporization rate at the interface. Since $j_{\gamma}$ is only non-zero at the interface cells, it is difficult to visualize. For better visualization, we have ``thickened'' the thin region where vaporization occurs by extrapolating the value of $j_{\gamma}$ outwards in the normal direction away from the interface for a distance of $0.2R_0$. Note that the ``thickening" of the region with non-zero $j_{\gamma}$ is only performed during the post-processing step for visualization purposes, and it does not influence the solution process. It is observed that the temperature gradient magnitude is significantly higher on the windward surface of the drop, where the local vaporization rate $j_\gamma$ is correspondingly larger. This observation is consistent with former studies \cite{renksizbulut_numerical_1983}. The vapor generated from the drop is constrained in the wake of the drop, which exhibits a lower temperature compared to that in the free stream. As a result, the leeward surface of the drop is exposed to lower-temperature vapor, and therefore, the temperature gradient and vaporization rate are lower than those on the windward surface.

\begin{figure}[tbp]
	\begin{center}
		\includegraphics [width=1.0\textwidth]{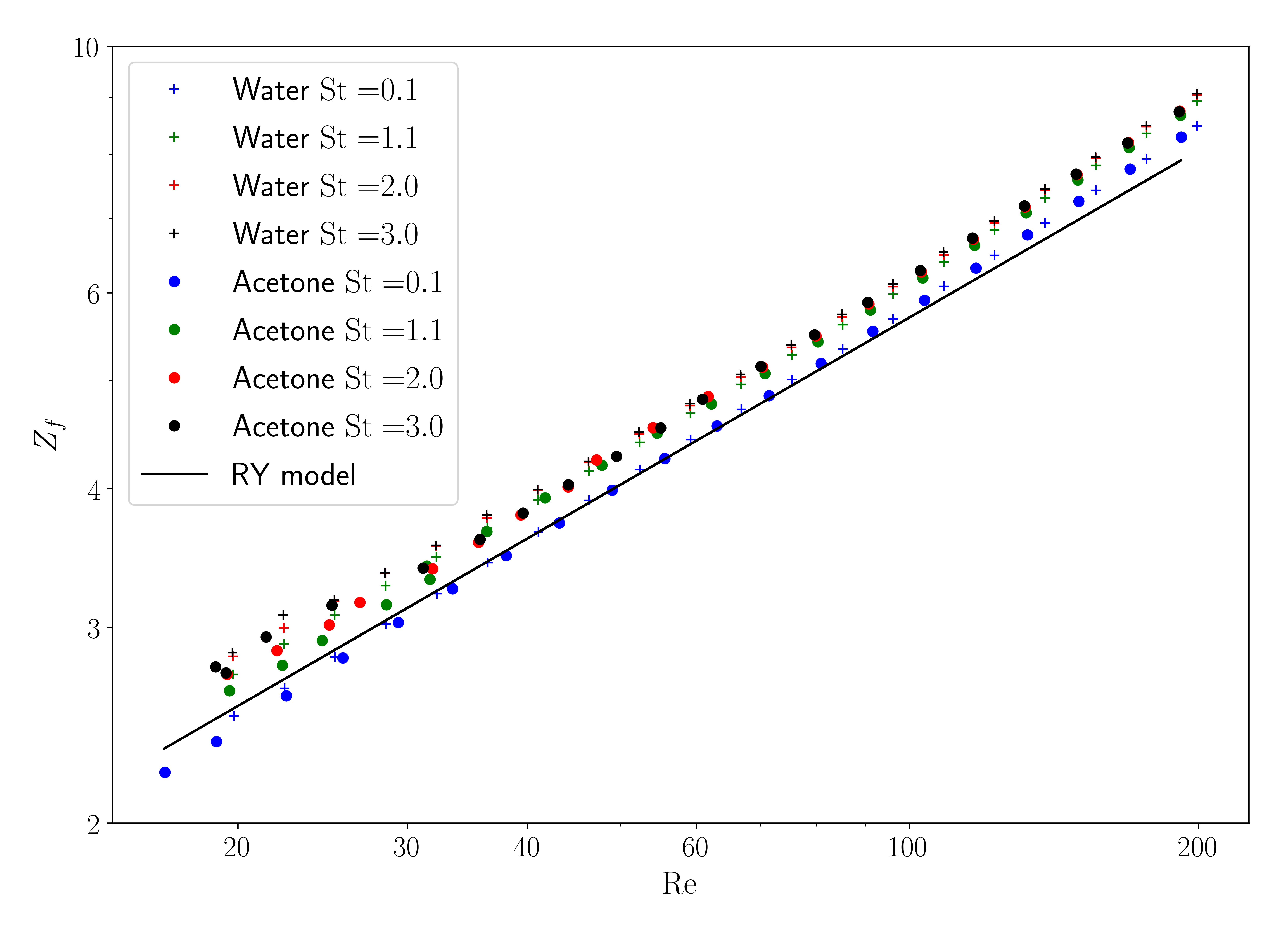}
	\end{center}
	\caption{Simulation results for different $\text{Re}$ between 20 and 200, at various St ($\text{St}=0.1$, $0.75$, $1.4$, $2.05$) for water and acetone (Table \ref{tab:properties}), compared with the empirical correlation ($Z_f$ is defined in Eq.~\eqref{eq:Zf}).}
	\label{fig:empir_vs_sim} 
\end{figure}

In the simulation, the volume of the liquid drop $V_l$ can be computed from the $f$ field at a given time. Based on $V_l$, the rate of drop volume change $\dot{V}_l=dV_l/dt$ can be evaluated. The dimensionless drop vaporization rate is $\dot{V}_l^*=dV_l^*/dt^* = \dot{V}_l \tau_c/(\pi D_0^3/6)$. As the drop vaporization is driven by heat transfer, $\dot{V}_l$ is generally a function of $\text{Nu}$ and $\text{St}$, as shown in \ref{app:V_dot}. Therefore, the $\dot{V}_l$ measured in the quasi-steady state can be used to calculate the corresponding $\text{Nu}$. To validate the present simulation methods, the simulation results of $\text{Nu}$ are compared with the RY empirical correlation in Fig.~\ref{fig:empir_vs_sim}. The RY empirical correlation is for quasi-steady vaporization for a spherical drop with constant volume, for which $\text{Nu}$ is expressed as a function of $\text{Re}$, $\text{Pr}$, and $\text{St}$ \citep{renksizbulut_experimental_1983}, 
\begin{align}
		\text{Nu}_{s} = [0.57 \text{Re}^{{1}/{2}}\text{Pr}^{{1}/{3}} + 2](1 + \text{St})^{-0.7}. 
 \label{eq:empirical}
\end{align}
The RY correlation is valid to predict the vaporization rate of a spherical drop for $25<\text{Re}<2000$, $0.07<\text{St}<2.79$, and $0.7<\text{Pr}<1$. The subscript $s$ is used to denote properties related to a spherical drop. Instead of plotting $\text{Nu}$ directly, the parameter $Z_f$ is used in Fig.~\ref{fig:empir_vs_sim}, following previous studies \cite{renksizbulut_experimental_1983, renksizbulut_numerical_1983}. The definition of $Z_f$ is given as
\begin{align}
Z_f = [\text{Nu} (1 + \text{St})^{0.7} - 2]\text{Pr}^{-{1}/{3}}.
\label{eq:Zf}
\end{align}
According to the RY correlation for a spherical drop (Eq.\eqref{eq:empirical}), $Z_{f}$ is only a function of $\text{Re}$:
\begin{align}
Z_{f} = 0.57 \text{Re}^{{1}/{2}},
\label{eq:Zf_exp}
\end{align}
The simulation results shown here correspond to a mesh resolution of $L=12$, which has been confirmed to be mesh-independent in the grid-refinement study results in \ref{app:drop_converge}. It is observed that the simulation results match well with the empirical correlation for the whole range of $\text{Re}$ considered. According to experimental data \cite{renksizbulut_experimental_1983, renksizbulut_numerical_1983}, $Z_f$ is independent of $\text{St}$. The simulation results are consistent with that empirical observation, showing that $Z_f$ for different $\text{St}$ are very similar.

The small discrepancies between the simulation results and the empirical correlation can be attributed to the minor deformation in the simulations. In the experiment, the drop was constrained to be perfectly spherical, which was not the case in the simulations. Nevertheless, the generally good agreement observed here validates the present methods in simulating the vaporization of a deformable drop.

\subsection{Vaporization of drops with low Weber numbers}
\label{section:low_We}
With the validated simulation approach, we investigate the vaporization of drops with finite Weber numbers. Only acetone drops are considered for the remainder of the paper. This section focuses on low $\text{We}_0$ values, ranging from 0.5 to 10.5, for which the drop shape oscillates in time and breakup typically does not occur ($\text{We}_0<\text{We}_{cr}$).

\begin{figure}[tbp]
	\begin{center}
		\includegraphics [width=1.0\textwidth]{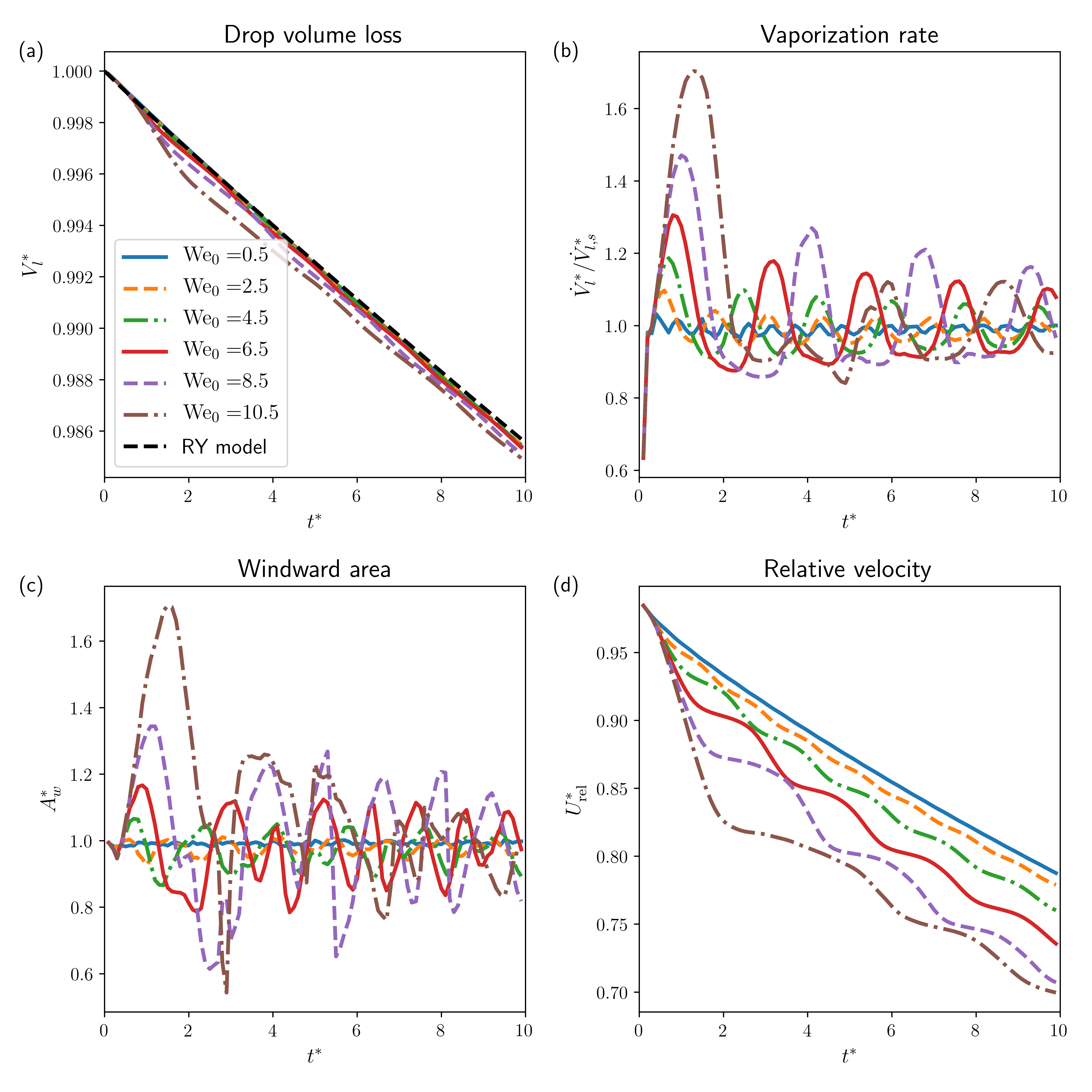}
	\end{center}
	\caption{Temporal evolutions of (a) drop volume, (b) the normalized vaporization rate $\dot{V}_{l}/\dot{V}_{l,s}$, (c) windward surface area of the drop, and (d) the relative velocity of the drop for $0.5\le \text{We}_0\le 10.5$, while $\text{Re}_0=500$ and $\text{St}=0.1$.}
	\label{fig:evap_vary_low_We} 
\end{figure}

In the simulations, $\text{St}=0.1$ and $\text{Re}_0=500$ are fixed to enable a parametric study of the effect of $\text{We}_0$ only. The domain size is increased to $l=16D_0$ to ensure that the drop remains far away from the domain boundary during the simulation time. The maximum refinement level is increased to $L=14$ (equivalent to $\Delta x = D_0/1024$) to accurately resolve the higher-level drop deformation. Due to vaporization, the drop volume $V_l$ decreases with time (see Fig.~\ref{fig:evap_vary_low_We}(a)). As a consequence of the drag on the drop, $U_\text{rel}$ and $\text{Re}$ also decrease with time (see Fig. \ref{fig:evap_vary_low_We}(d)). Since both $\text{Re}$ and drop deformation contribute to the variation of the drop vaporization rate, we normalize the instantaneous vaporization rate $\dot{V}_l^*$ by its counterpart for a spherical drop with the same $\text{Re}$, \ie, $\dot{V}_{l,s}^*$, to examine the effect of drop deformation on the vaporization rate. The expression of $\dot{V}_{l,s}^*$ is given in Eq.~\eqref{eq:volume_evap_rate_APPEN3}, and the derivation details can be found in \ref{app:V_dot}. The results of $\dot{V}_l^*/\dot{V}_{l,s}^*$ for different $\text{We}_0$ are shown in Fig.~\ref{fig:evap_vary_low_We}(b). Since the drop is initially spherical, $\dot{V}_l^*/\dot{V}_{l,s}^*$ approaches about 1 after the short transition due to thermal boundary layer development. Although $V_l^*$ decreases monotonically over time, $\dot{V}_l^*/\dot{V}_{l,s}^*$ oscillates as time elapses, which is related to the drop shape oscillation. Generally, the oscillation amplitude of $\dot{V}_l^*/\dot{V}_{l,s}^*$ is higher for cases with larger $\text{We}_0$.

It has been shown previously in Fig.~\ref{fig:moving_drop_temperature} that the local vaporization rate is higher on the windward surface of the drop. To measure the windward surface area $A_w$ in the simulations, we have summed the area of VOF reconstructed interfaces for which the normal vector $\mathbf{n}$ is facing in the upstream direction (\ie, $n_x<0$). The windward surface area of the drop is normalized also by its spherical counterpart, $A_w^* = {A_w}/{A_{w,s}}$, where $A_{w,s}=\pi D^2/2$. The temporal evolutions of $A_w^*$ for different $\text{We}_0$ are shown in Fig.~\ref{fig:evap_vary_low_We}(c) and it is clear that $A_w^*$ oscillates in time due to the drop shape oscillation. The time oscillation of $A_w^*$ will lead to a large area of high local vaporization rate, the overall drop vaporization rate will thus oscillate. The similar time evolutions between $\dot{V}_l^*/\dot{V}_{l,s}^*$ and $A_w^*$ are clear evident. 

Finally, it is worth noting that, the shape oscillation also contributes to the time oscillation of drag. As a result, the decrease of $U_\text{rel}^*=U_\text{rel}/U_\infty$ is not monotonic, see Fig.~\ref{fig:evap_vary_low_We}(d). In particular, the rapid rise of $A_w^*$ for large $\text{We}$ at an early time results in a significant increase of drag, which in turn induces a rapid decrease in relative velocity. A reduced relative velocity will result in a decrease of $\text{Re}$, which will in turn reduce the drop vaporization rate (see Eq.~\eqref{eq:empirical}). Nevertheless, the effect of $\text{Re}$ on the drop vaporization rate is generally small, and the main factor that drives the drop vaporization rate oscillation is the drop deformation and the resulting oscillation of $A_w^*$.  

\subsection{Vaporization of drops with moderate Weber numbers}
\label{section:moderate_We}
\begin{figure}[tbp]
	\begin{center}
		\includegraphics [width=0.8\textwidth]{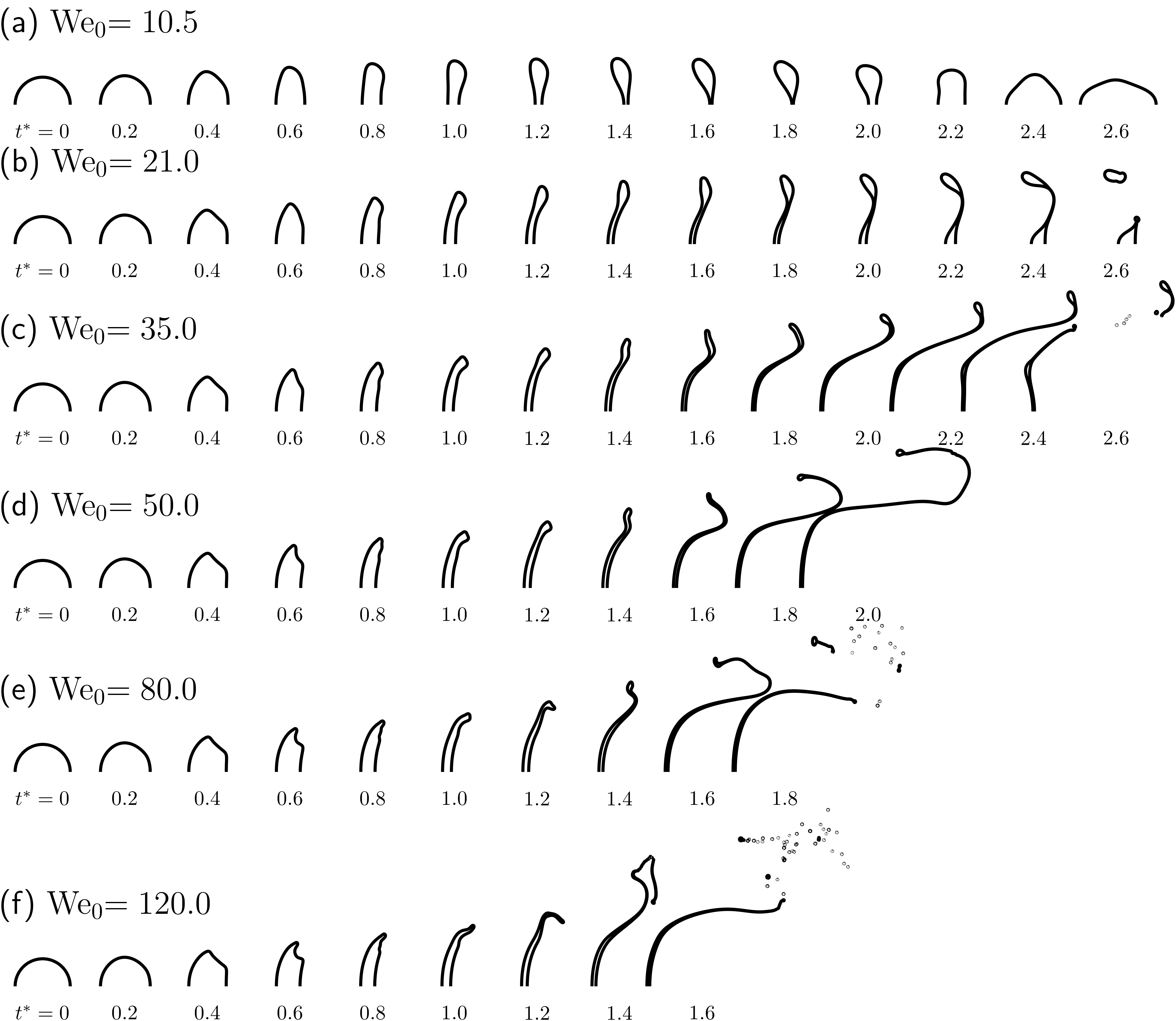}
	\end{center}
	\caption{The temporal evolutions of the drop shape for various Weber numbers: (a) $\text{We}_0=10.5$, (b) $21$, (c) $35$, (d) $50$, (e) $80$, and (f) $120$.}
	\label{fig:drop_shapes} 
\end{figure}

When $\text{We}_0$ is increased to a moderate range, \ie, $\text{We}_0=21$ to $120$, the drop deformation becomes more significant, and eventually a topological change will occur. Similar to the previous section, $\text{St=0.1}$ and $\text{Re}_0=500$ are fixed while $\text{We}_0$ is varied. The temporal evolutions of the drop shapes for different $\text{We}_0$ are shown in Fig.~\ref{fig:drop_shapes}. It is observed that drops with moderate $\text{We}_0$ show a substantially different morphological evolution, compared to those with low $\text{We}_0$. According to the simulation results, the critical Weber number for the acetone drops considered is about 21. The drop with moderate $\text{We}_0$ first deforms to a disk. As the lateral edge bends downstream, the disk evolves into a backward bag, with the opening facing downstream.  A rim forms at the edge of the bag \cite{marcotte_density_2019}. In the case of low-Oh liquids such as water and acetone, the dynamics of the rim are dominated by capillary effects and are well represented by the Taylor-Culick theory \cite{deka_revisiting_2020, agbaglah_longitudinal_2013}. For Taylor-Culick rims, the minimum thickness of the sheet occurs at the neck of the edge rim \cite{marcotte_density_2019}. The liquid sheet with a small thickness moves faster downstream, leading to the formation of a forward (upstream-facing) ring bag. In most cases, the inflation and rupture of the ring bag signal the onset of drop breakup. 
The breakup behavior of the acetone drops for different $\text{We}_0$, as is described above, is similar to previous numerical studies for similar density ratios ($\eta$) \cite{marcotte_density_2019, jain_secondary_2019}. In the present study, we focus on drop vaporization before breakup occurs. Capturing the complete disintegration of the drop and the vaporization of children drops generated will require significantly longer 3D simulations, which will be relegated to future works.

As $\text{We}_0$ is increased further, the backward bag and forward ring bag will appear earlier. The lateral radius of the bag also increases faster when $\text{We}_0$ increases. It is also worth noting that for $\text{We}=120$, the rapid lateral expansion of the drop will suppress the formation of the rim and the edge of the drop is bent downstream by the strong shear and the entrainment of the separated flow. These are typical features that indicate the drop breakup mode is in transition to the shear mode \cite{marcotte_density_2019}.  

\begin{figure}[tbp]
	\begin{center}
		\includegraphics [width=1.0\textwidth]{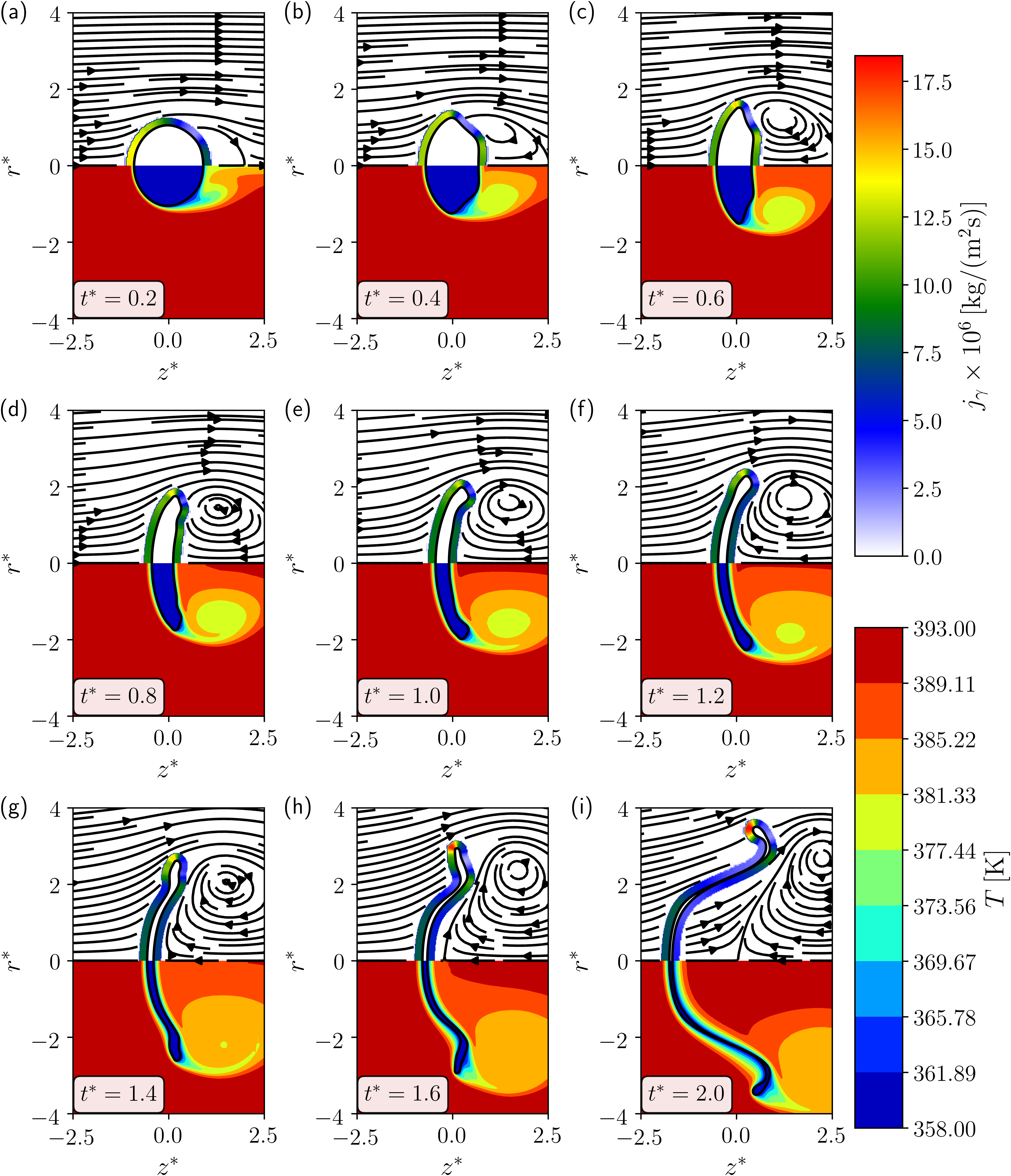}
	\end{center}
    	\caption{The temperature, streamlines, and vaporization rate ($j_{\gamma}$) during  the deformation of a drop where $\text{We}_0=35$. The streamlines are for the gas velocity field in the drop reference frame.}
	\label{fig:drop_shape_const_rel_vel} 
\end{figure}

Figure \ref{fig:drop_shape_const_rel_vel} shows the temporal evolutions of the streamlines, vaporization rate, and temperature for drops with $\text{We}_0=35$. Here the streamlines are for the velocity field in the drop reference frame. At the early time $t^*=0.2$, the drop shape, the gas flow, and the temperature distribution are all quite similar to those for spherical drops (Fig.~\ref{fig:moving_drop_temperature}). Therefore, the value of $j_\gamma$ on the windward surface is significantly higher than that on the leeward surface. As time progresses, the drop deforms into a disk and a backward bag, increasing the windward area of the drop and the size of the wake (Figure \ref{fig:drop_shape_const_rel_vel}(b)-(f)). An interesting difference from the spherical drop is that the high-temperature vapor in the free stream is entrained into the wake along the axis of symmetry. As a result, the vaporization rate near the leeward pole increases. It can be observed that the values of $j_\gamma$ at the windward and leeward poles of the drop are similar at $t^*=0.8$. As the drop continues to deform and the rim develops at the lateral edge, it is observed that the maximum temperature gradient and local vaporization rate appear on the windward side of the rim, see $t^*=1.6$.

\begin{figure}[tbp]
	\begin{center}
		\includegraphics [width=1.0\textwidth]{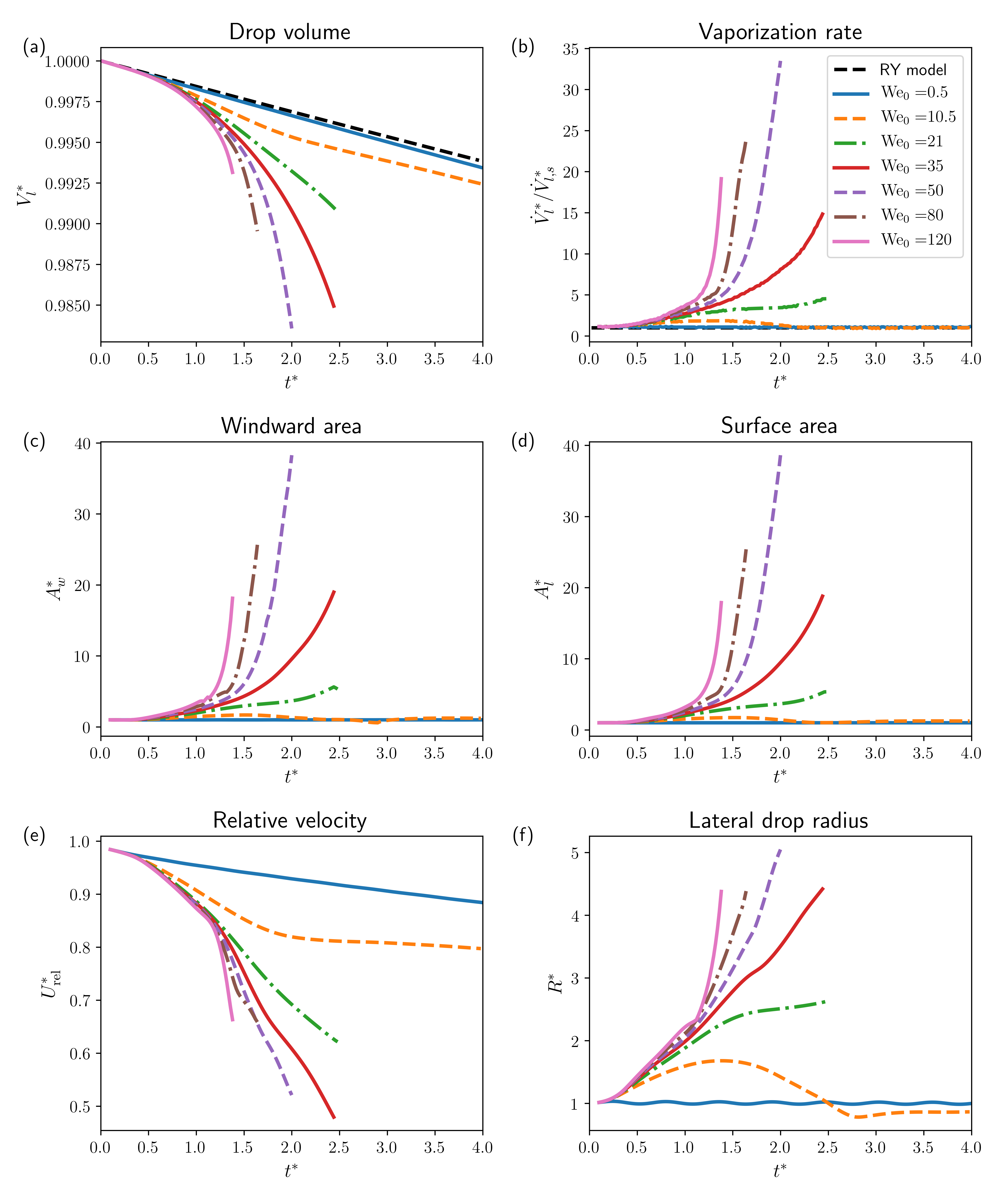}
	\end{center}
	\caption{Time evolutions of (a) the drop volume, (b) the vaporization rate, (c) the windward area, (d) the total surface area, (e) the relative velocity, and (f) the radial deformation of the drop until breakup occurs for various Weber numbers: $\text{We}_0=0.5$, $10.5$, $21$, $35$, $50$, $80$, and $120$ at $L=14$.}
	\label{fig:drop_evap_We} 
\end{figure}

The temporal evolutions for $V_l^*$,  $\dot{V}_l^*/\dot{V}_{l,s}^*$, $A_w^*$, and $U_\text{rel}^*$ are shown in Fig.~\ref{fig:drop_evap_We}. Since the local vaporization rate $j_\gamma$ on the leeward surface can also be as significant as that on the windward surface, the total surface area of the drop, $A_l$, is also plotted. Similarly to $A^*_w$, we have also normalized $A_l$ also by its spherical counterpart, \ie, $A_{l}^*=A_l/(\pi D^2)$. It is observed that  $A_l^*$  is very similar to $A_w^*$ for all time. The reason $A_w^* \approx A_l^*$ is that when the drop exhibits a spheroid shape at an early time or a bag shape at a later time, the windward and leeward surface areas are similar. In such a case, we will retain the use of $A^*_w$ to characterize the deformation of the drop and its impact on the rate of drop vaporization, although the local vaporization on the leeward surface is also of importance. 

The evolutions of $\dot{V}_l^*/\dot{V}_{l,s}^*$ for different $\text{We}_0$ are similar to that of a spherical drop until around $t^*=0.2$ as the drop deformation is still mild at this stage. However, as the drop continues to deform into a bag shape after $t^*>0.5$, $A_w^*$ increases much more rapidly due to bag inflation. The rapid increase in $A_w^*$ leads to a corresponding increase in the rate of drop vaporization, $\dot{V}_l^*/\dot{V}_{l,s}^*$. Additionally, the larger $A_w^*$ contributes to a higher drag, causing $U_\text{rel}^*$ to increase rapidly as well. Since the rate of increase in $A_w^*$ generally increases with $\text{We}_0$, the evolutions of $\dot{V}_l^*/\dot{V}_{l,s}^*$ and $U_\text{rel}^*$ for different $\text{We}_0$ become more distinct from each other.

\begin{figure}[tbp]
	\begin{center}
		\includegraphics [width=1.0\textwidth]{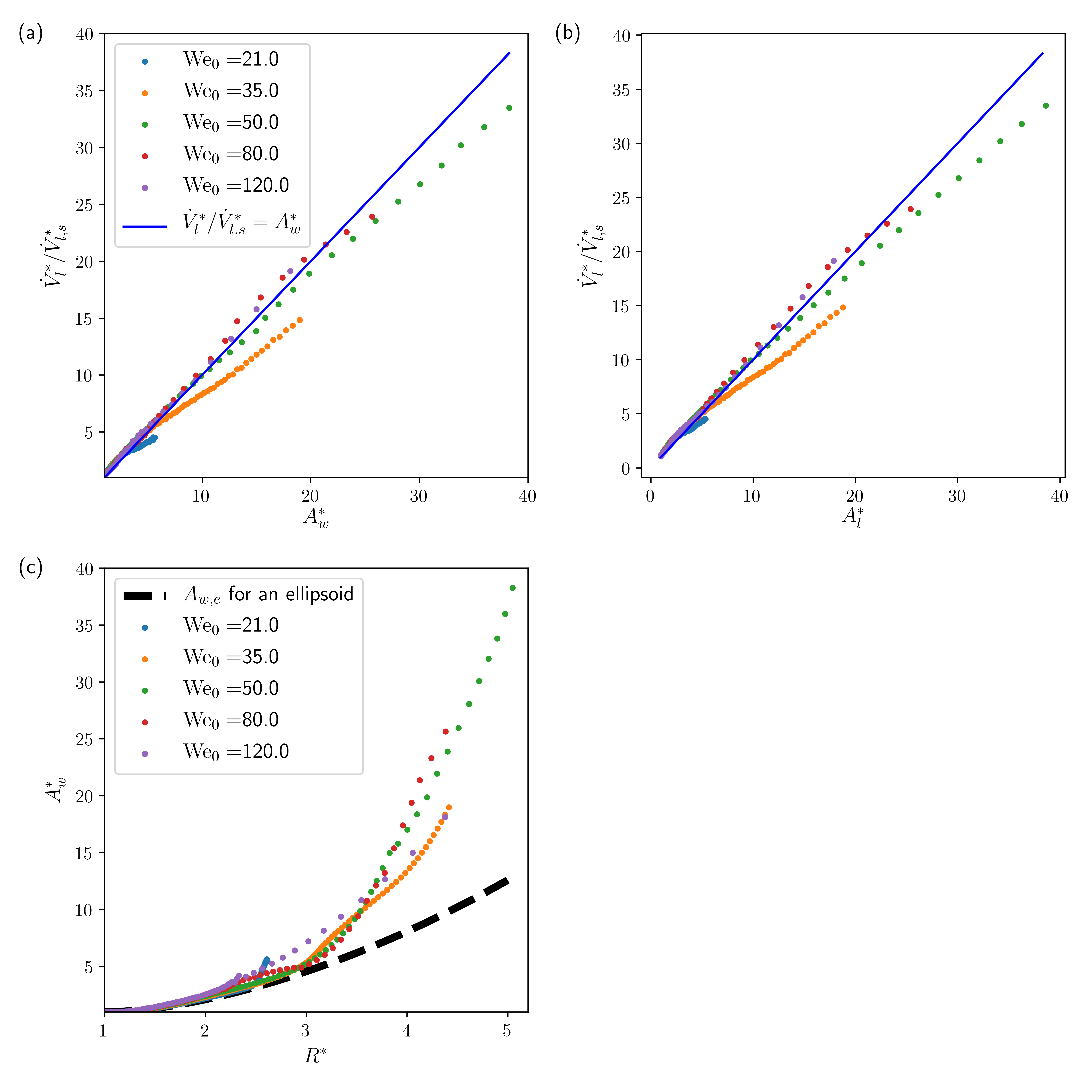}
	\end{center}
	\caption{Drop vaporization rate as a function of the (a) windward area and (b) surface area for moderate Weber numbers, and (c) drop windward area as a function of lateral radius. Note that $A_{w,e}$ is defined in Eq.~\eqref{eq:Af_approx}}
	\label{fig:vap_correlation_moderate_We} 
\end{figure}

Unlike drops with low $\text{We}_0$, drops for $\text{We} \geq 21$ do not experience any shape oscillation. As a result, the magnitudes of $\dot{V}_l^*/\dot{V}_{l,s}^*$, $A_w^*$, and $U_\text{rel}^*$ all increase monotonically over time. It is observed that the evolutions of $\dot{V}_l^*/\dot{V}_{l,s}^*$ and $A_w^*$ are again similar. To better depict the relation between the two, $\dot{V}_l^*/\dot{V}_{l,s}^*$ is plotted as a function of $A_w^*$ in Fig.~\ref{fig:vap_correlation_moderate_We}(a). It is clearly seen that $\dot{V}_l^*/\dot{V}_{l,s}^*$ increases approximately linearly over time, and the slope is approximately one. The high correlation between $\dot{V}_l^*/\dot{V}_{l,s}^*$ and $A_w^*$ indicates that the drop vaporization rate is mainly influenced by the instantaneous shape and that the temporal shape evolution has a secondary effect. This observation is important for the development of a vaporization model for deformable drops, which will be presented later.

In previous studies of drop aerobreakup \cite{guildenbecher_secondary_2009}, the evolution of the lateral drop radius $R^*=2R/D_0$ is used to characterize the drop deformation. The lateral radius of the drop $R$ is defined as the maximum interfacial position from the axis of symmetry. The results of $R^*$ for different $\text{We}_0$ are shown in Fig.~\ref{fig:drop_evap_We}(f). At early time, the shape can be well approximated as an ellipsoid. The results of $A_w^*$ are plotted as a function of $R^*$ in Fig.~\ref{fig:vap_correlation_moderate_We}(c), and it can be observed that the simulation results for different $\text{We}_0$ are quite similar to the estimate assuming the drop shape to be ellipsoidal. After about $R^*=2.5$, the forward ring bag starts to form and inflates, then a significant deviation from the ellipsoidal estimate is observed. Deviations for the temporal evolutions of $R^*$ for different $\text{We}_0$ can also be seen in Fig.~\ref{fig:drop_evap_We}(f) for $R^*\ge2.5$. This is due to the fact that the formation and development of the forward ring bag are very sensitive to $\text{We}_0$. 

\subsection{Effect of $\text{Re}$}
\label{section:influence_of_Re}
The previous sections have presented results for a fixed value of $\text{Re}_0=500$. To investigate the influence of $\text{Re}_0$ on the drop vaporization rate and its relation with drop deformation, a parametric study of $\text{Re}$ is performed in this section. The value of $\text{Re}_0$ is varied from 50 to 500 by changing $U_\infty$ and $\sigma$ simultaneously, while keeping the values of $\text{We}_0$ and $\text{St}$ constant at 35 and 0.1, respectively.

\begin{figure}[tbp]
	\begin{center}
		\includegraphics [width=1.0\textwidth]{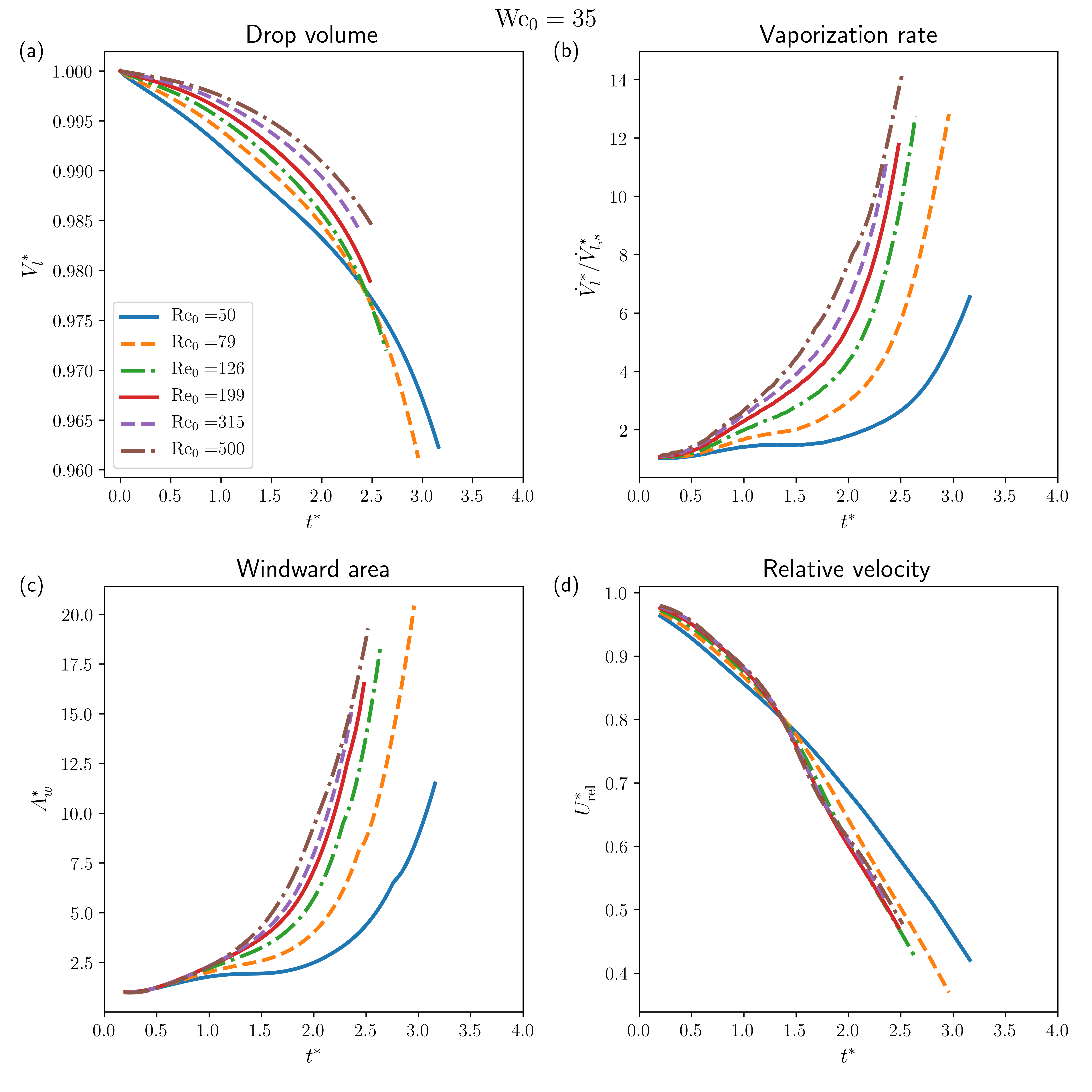}
	\end{center}
	\caption{Temporal evolutions of (a) the drop volume, (b) the normalized vaporization rate $\dot{V}_l^*/\dot{V}_{l,s}^*$, (c) the windward area, and (d) the relative velocity of the drop until breakup occurs for different $\text{Re}_0$ while $\text{We}_0=35$ is fixed.}
	\label{fig:drop_evap_We35_RE} 
\end{figure}

The temporal evolution of the drop volume $V_l^*$ is shown in Fig.~\ref{fig:drop_evap_We35_RE}(a). It is observed that the normalized drop volume $V_l^*$ initially decreases faster for lower $\text{Re}_0$. This is because the initial drop vaporization rate is similar to that of a spherical drop, $\dot{V}_{l,s}^*$, which decreases with $\text{Re}_0$ as shown in Eq.~\eqref{eq:volume_evap_rate_APPEN3}. Nevertheless, as time evolves and the drop deforms, the vaporization rate for drops with higher $\text{Re}_0$ increases faster; see Fig.~\ref{fig:drop_evap_We35_RE}(b). For example, the volume $V_l^*$ of the drop with $\text{Re}_0=50$ initially decreases faster than that for $\text{Re}_0=79$, indicating that the former exhibits a higher vaporization rate. However, the drop with $\text{Re}_0=79$ deforms faster over time; consequently, its vaporization rate also increases more rapidly. As a consequence, the $V_l^*$ of the drop with $\text{Re}_0=79$ falls below that of the drop with $\text{Re}_0=50$ at approximately $t^*=2.5$.

\begin{figure}[tbp]
	\begin{center}
		\includegraphics [width=1.0\textwidth]{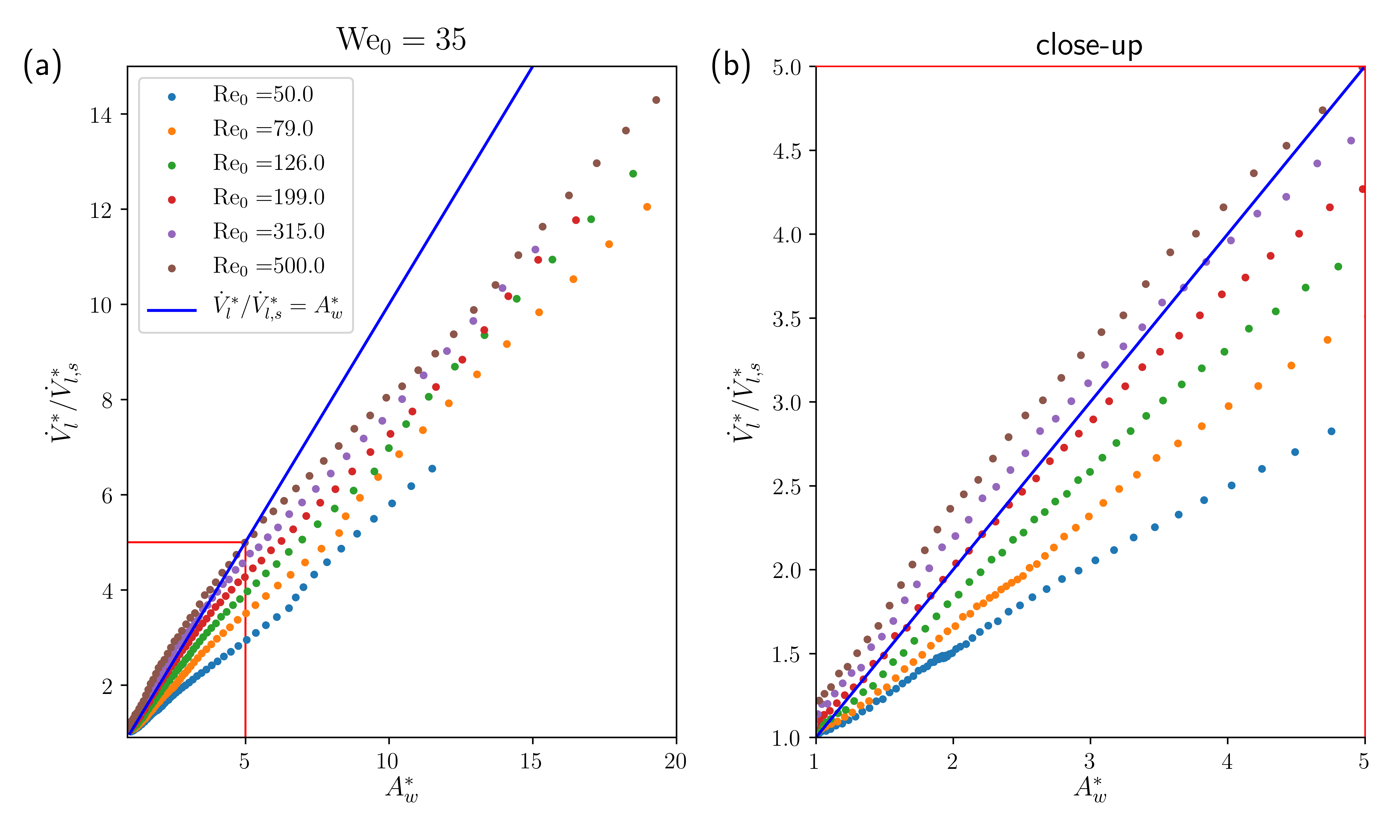}
	\end{center}
	\caption{(a) Correlation between vaporization rate and the windward area of the drop for $\text{We}=35$ at various $\text{Re}_0$: $\text{Re}_0=50$, $79$, $126$, $199$, $315$, and $500$. (b) Close-up of results for $A_w^*<5$.}
	\label{fig:vap_correlation_moderate_We35_RE} 
\end{figure}

Generally, the normalized vaporization rate increases more rapidly as $\text{Re}_0$ increases, which is consistent with the trend for $A_w^*$, as shown in Fig.~\ref{fig:vap_correlation_moderate_We35_RE}(c). A change in $\text{Re}_0$ will influence the drag on the drop and the interfacial Rayleigh-Taylor instability (RTI) that is triggered by the drag-induced acceleration of the drop. Therefore, $\text{Re}_0$ has an impact on the drop deformation and the resulting windward surface area. For the results shown here, the change in the drop deformation may also be related to the variation of $\text{Oh}$ as we have varied $\sigma$ to change $\text{Re}_0$ while keeping $\text{We}_0$ fixed. Here, $\text{Oh}$ decreases as $\text{Re}$ increases. Nevertheless, since $\text{Oh}$ is generally small, the effect of its variation is expected to be secondary.

For drops with $\text{Re}_0=500$ shown in previous sections, it has been observed that the normalized vaporization rate $\dot{V}_l^*/\dot{V}_{l,s}^*$ is closely related to the windward surface area. The results for $\dot{V}_l^*/\dot{V}_{l,s}^*$ for different $\text{Re}_0$  are plotted as a function of $A_w^*$ in Fig.~\ref{fig:vap_correlation_moderate_We35_RE}. It is observed that $\dot{V}_l^*/\dot{V}_{l,s}^*$ increases monotonically with $A_w^*$ for all $\text{Re}_0$. For $A_w^*\lesssim 5$, \ie, before the ring bag inflates, the variation is approximately linear, see Fig.~\ref{fig:vap_correlation_moderate_We35_RE}(b). The increasing rate of $\dot{V}_l^*/\dot{V}_{l,s}^*$ over  $A_w^*$ is reduced as  $\text{Re}_0$ decreases. For example, the slope for $\text{Re}_0=500$ is about 1, while that for $\text{Re}_0=50$ decreases to about 0.5. 

\begin{figure}[tbp]
	\begin{center}
		\includegraphics [width=1.0\textwidth]{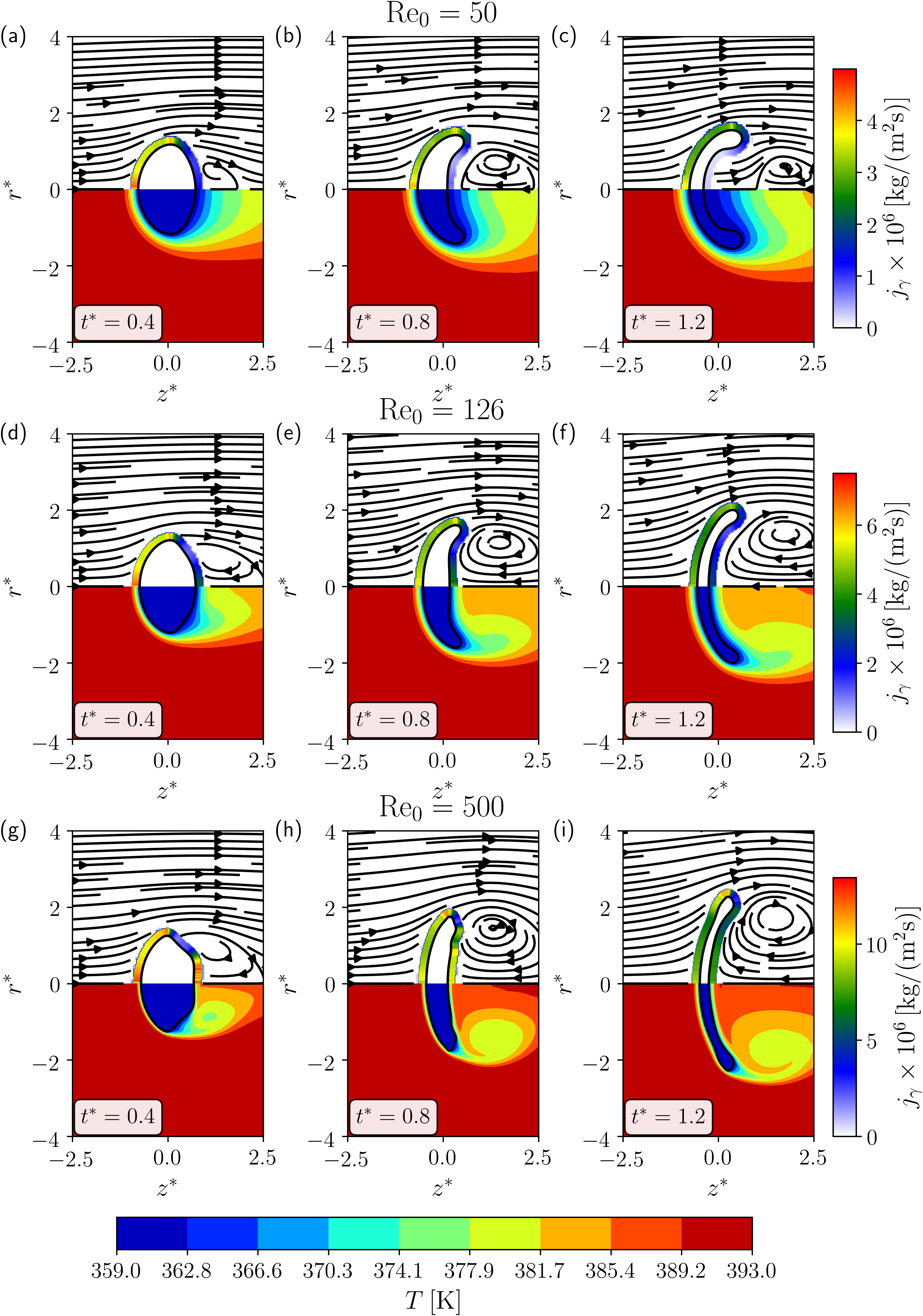}
	\end{center}
    	\caption{The vaporization rate ($j_{\gamma}$), streamlines, and temperature field for (a)-(c)  $\text{Re}_0=50$, (d)-(f) $\text{Re}_0=126$, and (g)-(i) $\text{Re}_0=500$ where $\text{We}_0=35$.}
	\label{fig:j_dot_Re} 
\end{figure}

The effect of $\text{Re}_0$ on the variation of $\dot{V}_l^*/\dot{V}_{l,s}^*$ over $A_w^*$ is related to drop deformation and heat transfer in the wake. For drops with low $\text{Re}_0$ ($\text{Re}_0=50$), thermal diffusion dominates, resulting in low gas temperature and gradient on the leeward side of the drop. As a result, most of the vaporization occurs on the windward surface due to the higher temperature gradient, as shown in Figs.~\ref{fig:j_dot_Re}(a)-(c). As $\text{Re}_0$ increases, stronger convection entrains hot vapor from the free stream in the wake, resulting in increased temperature and gradient on the leeward surface of the drop, leading to an increase in the local vaporization rate $j_\gamma$, as shown in Figs.~\ref{fig:j_dot_Re}(d)-(f). When $\text{Re}_0$ reaches 500, the vaporization rates on the windward and leeward surfaces become comparable. This explains why $\dot{V}_l^*/\dot{V}_{l,s}^*$ for $\text{Re}_0=500$ is approximately twice that for $\text{Re}_0=50$ for the same $A_w^*$, as effective vaporization occurs on the surface about $2A_w^*$ for the former. The enhanced convective heat transfer for higher $\text{Re}_0$ is also related to the more significant deformation, which in turn changes the flow separation and wake structure.

\subsection{Results for full 3D simulations}
\label{section:3D_section}
\begin{figure}[tbp]
	\begin{center}
		\includegraphics [width=0.7\columnwidth]{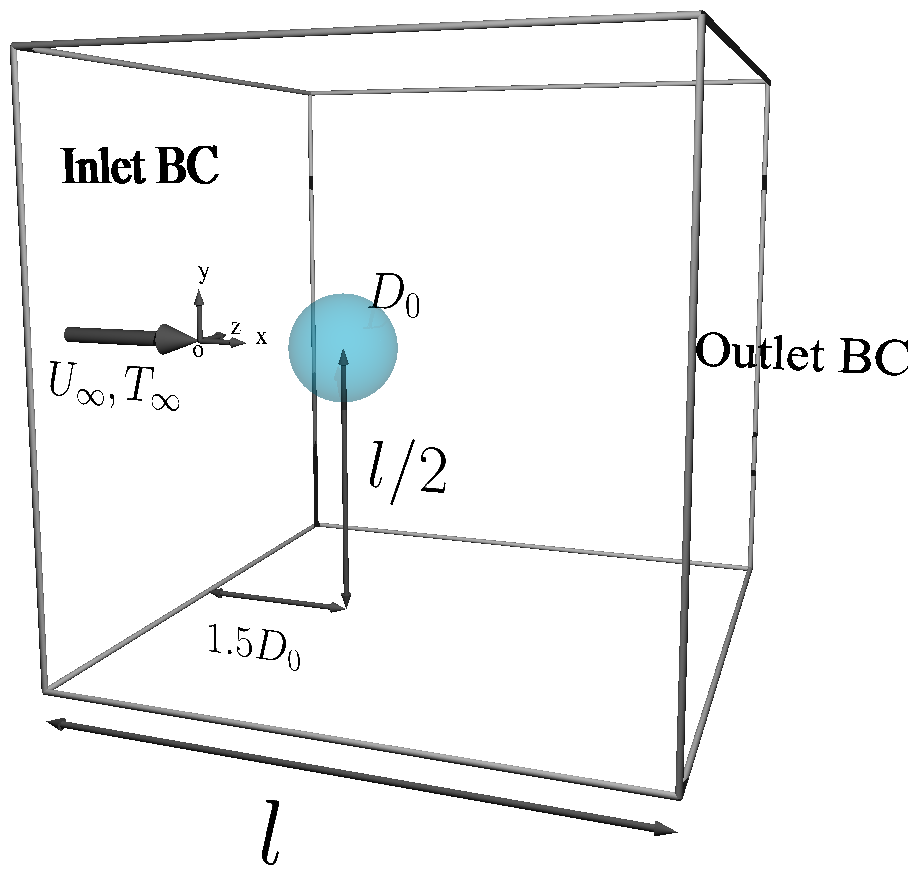}
	\end{center}
	\caption{Schematic of the 3D cubic computational domain for the aerobreakup of an acetone drop in a uniform high-temperature vapor stream.}
	\label{fig:3D_moving_droplet_diagram} 
\end{figure}

Fully 3D simulations for representative cases of acetone drops at different Weber and Reynolds numbers are conducted to verify if the 2D axisymmetric simulation results presented above are sufficient to capture the vaporization and drop deformation. A schematic of the 3D computational domain is given in Fig.~\ref{fig:3D_moving_droplet_diagram}. The maximum level of refinement is $L=12$, resulting in a minimum cell dimension $\Delta x=D_0/256$. The total number of octree cells reaches a maximum of about 80 million (equivalent to 69 billion uniform Cartesian cells). 

\begin{figure}[tbp]
	\begin{center}
		\includegraphics [width=1.0\textwidth]{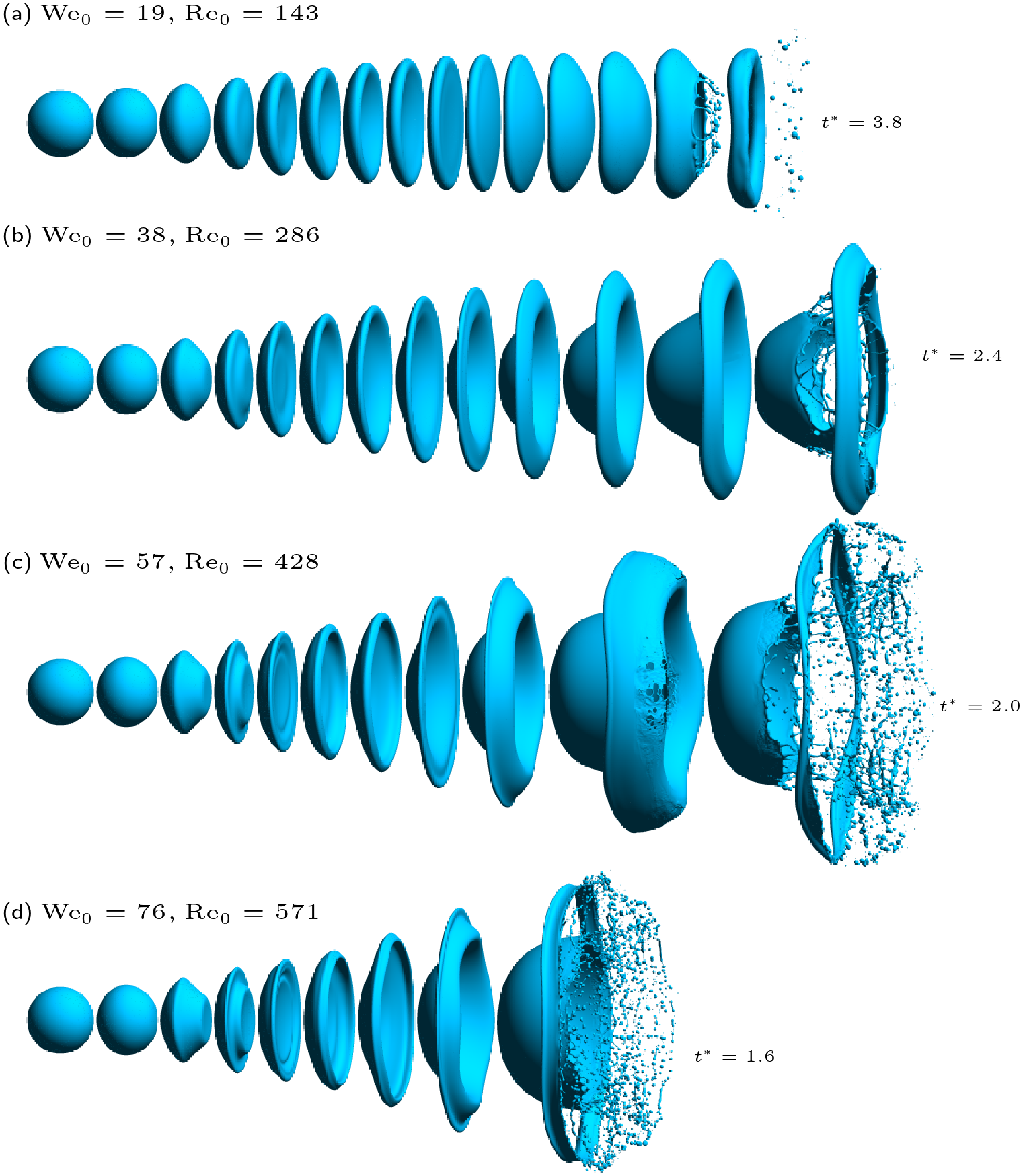}
	\end{center}
	\caption{The drop evolution at time increments of $t^*=0.2$ for moderate Weber numbers and at various Reynolds numbers: (a) $[\text{We}_0, \text{Re}_0]=[19,143]$, (b) $[\text{We}_0, \text{Re}_0]=[38,286]$, (c) $[\text{We}_0, \text{Re}_0]=[57,428]$, and (d) $[\text{We}_0, \text{Re}_0]=[76,571]$.}
	\label{fig:3D_drop_evolution} 
\end{figure}

The four cases considered here are (1) $[\text{We}_0,\text{Re}_0]=[19,143]$, (2) $[\text{We}_0,\text{Re}_0]=[38,286]$, (3) $[\text{We}_0,\text{Re}_0]=[57,428]$, and (4) $[\text{We}_0,\text{Re}_0]=[76,571]$, while $\text{St}$ is kept constant at 0.1. The Ohnesorge numbers for the four cases are $\text{Oh}=0.05$, 0.035, 0.029, and 0.025, respectively. The morphological evolutions are presented in Fig.~\ref{fig:3D_drop_evolution}. For $\text{We}_0=19$, the drop first deforms to a small backward bag. But the backward bag does not continue to grow in length, instead is flattened back to a thin disk. Then the center of the disk moves faster, forming a forward bag. Eventually, the forward bag breaks in the center, leaving an unbroken liquid ring, see Fig.~\ref{fig:3D_drop_evolution}(a). This morphological evolution is typical for bag breakup for low-Oh drops near the critical $\text{We}_0$. When $\text{We}_0$ increases to larger than 38, a significant change in the drop morphology is observed. The backward bag will not be flattened back to a flat disk, instead, the length of the backward bag continues to increase in time. While the edge rim of the bag curves toward the upstream direction, a ring forward bag is formed between the backward bag at the center axis and the edge, as is annotated in Fig.~\ref{fig:drop_shapes}. The shape of the deformed drop before breakup is described as a multi-mode or rim-bag formation, which is also observed in former studies of drops at moderate Weber numbers and density ratios \cite{marcotte_density_2019, jain_secondary_2019}. Further increase of $\text{We}_0$ to 57 and 76 will result in the more rapid growth of the ring forward bag. The breakup typically starts when the sheet of the ring bag ruptures.  

\begin{figure}[tbp]
	\begin{center}
		\includegraphics [width=1.0\textwidth]{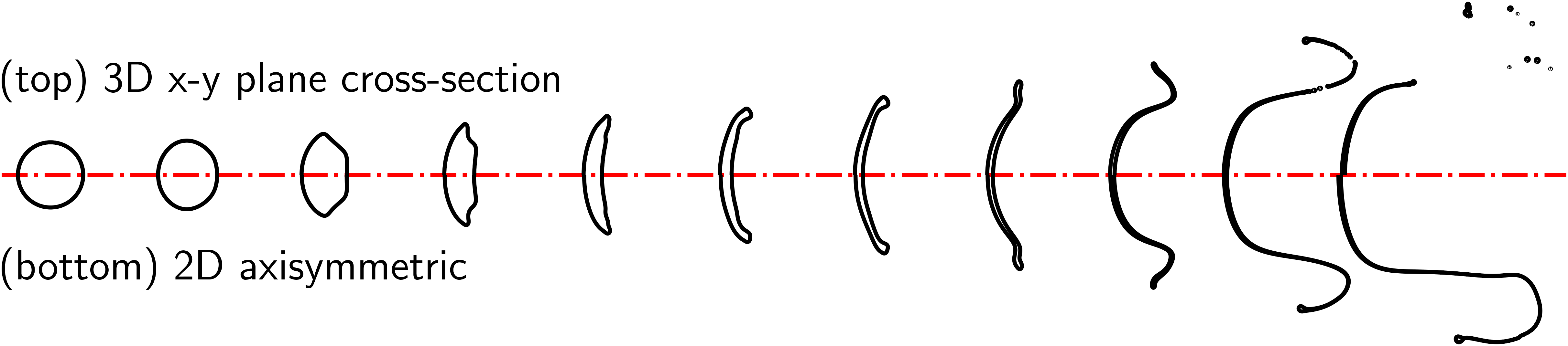}
	\end{center}
	\caption{Cross-section of the 3D drop evolution (top) compared to the 2D axisymmetric drop evolution (bottom) at time increments of $t^*=0.2$ for $[\text{We}_0, \text{Re}_0]=[76,571]$ }
	\label{fig:3D_vs_2D_shape} 
\end{figure}

Figure \ref{fig:3D_vs_2D_shape} directly compares the time evolutions of the drop shape obtained by the 3D and 2D axisymmetric simulations for $[\text{We}_0, \text{Re}_0]=[76,571]$. The 2D and 3D simulation results agree very well until the breakup occurs. The drops in the 3D simulation break earlier than their 2D counterparts, which is likely related to the additional stabilizing effect in 2D axisymmetric simulations, as well as the higher level of refinement achieved in the 2D simulations ($L=14$ and $L=12$ for the 2D and 3D simulations, respectively).

\begin{figure}[tbp]
	\begin{center}
		\includegraphics [width=1.0\textwidth]{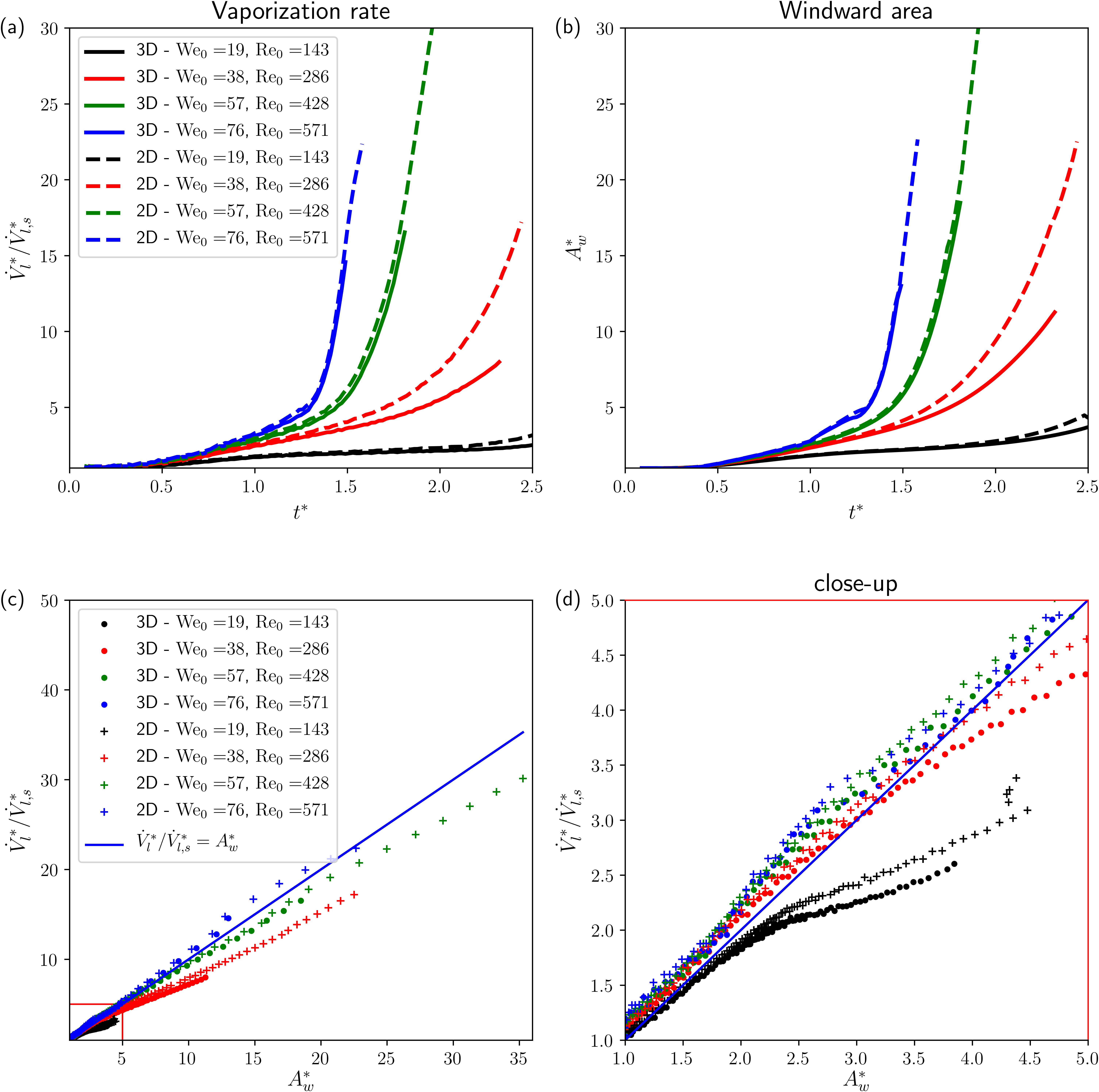}
	\end{center}
	\caption{The temporal evolutions of (a) the vaporization rate and (b) the windward area for moderate Weber numbers and at various Reynolds numbers: (1) $[\text{We}_0, \text{Re}_0]=[19,143]$, (2) $[\text{We}_0, \text{Re}_0]=[38,286]$, (3) $[\text{We}_0, \text{Re}_0]=[57,428]$, and (4) $[\text{We}_0, \text{Re}_0]=[76,571]$. The figure compares the 2D axisymmetric to 3D simulation results. (c)-(d) The vaporization rate is plotted as a function of the windward area where (d) is a close-up of (c).}
	\label{fig:3D_drop_vap_Af} 
\end{figure}

The evolutions of $\dot{V}_l^*/\dot{V}_{l,s}^*$ and $A_w^*$ for the 2D and 3D simulations at different $\text{We}_0$ and $\text{Re}_0$ are shown in Fig.~\ref{fig:3D_drop_vap_Af}. Again, the 2D and 3D results are very similar, confirming that 2D axisymmetric simulation is sufficient to capture the drop deformation and vaporization until breakup occurs. Furthermore, it can be observed from Figs.~\ref{fig:3D_drop_vap_Af}(c) and (d) that the 3D results for $\dot{V}_l^*/\dot{V}_{l,s}^*$ as a function of $A_w^*$ are also very similar to the 2D results. 

\section{Modeling the vaporization of a deformable drop}
\label{section:VTAB}
Based on the important observation in the simulation results, \ie, $\dot{V}_l^*/\dot{V}_{l,s}^*$ and  $A_w^*$ are closely related, a transient vaporization model for the deformable drop is proposed. Since $\dot{V}_l^*/\dot{V}_{l,s}^*$ increases approximately linearly over  $A_w^*$ and the slope is generally about unity for drops with larger $\text{Re}_0$, in the present model, we simply approximate 
\begin{align}
	\dot{V}_l^*(t) \approx \dot{V}_{l,s}^* A_w^*(t)\,, 
	\label{eq:volume_evap_rate_modified}
\end{align}
to capture the leading order effect of drop deformation on the vaporization rate. Since $A_w^*$ varies over time due to drop deformation, $\dot{V}_l^*$ also varies over time. The temporal evolution of $A_w^*$ is determined by the complicated drop deformation, which in turn depends on dimensionless parameters such as $\text{We}_0$, $\text{Re}_0$, and $\text{Oh}$ based on initial conditions and physical properties. A variety of drop deformation/breakup models have been developed  \cite{guildenbecher_secondary_2009}, though deviations are often observed between the model predictions, experimental results, and direct numerical simulation results. While improving the drop deformation model is beyond the scope of the present study, we use the commonly used Taylor Analogy Breakup (TAB) model to demonstrate how to incorporate the effect of drop deformation on the drop vaporization, resulting in the Vaporizing-TAB (V-TAB) model. Nevertheless, the model can be easily extended to incorporate other more sophisticated drop deformation models. 

The drop shape is approximated as an ellipsoid, the drop volume can thus be expressed as a function of the drop's lateral radius ($R$)
\begin{align}
V_l=\frac{4}{3}\pi a R^2\,
\end{align}
where $a$ is the length of the drop along the axis of symmetry. The windward surface area of an ellipsoid drop can be expressed as a function of $R$ and $V_l$
\begin{align}
A_{w,e} (R, V_l) \approx 2 \pi \left ( \frac{2(\frac{3V_{l}}{4\pi R})^{1.6} + R^{3.2}}{3} \right )^{1/1.6}\, .
\label{eq:Af_approx}
\end{align}
As shown in Fig.~\ref{fig:vap_correlation_moderate_We}(c), $A_{w,e}$ (Eq.~\eqref{eq:Af_approx}) is a good approximation of $A_w$ until about $R^*>3$, where the bag starts to inflate. The time evolution of $R$ is in turn estimated by treating the deforming drop as a forced and damped spring-mass system, according to the TAB model \cite{orourke_tab_1987}, \ie, 

\begin{align}
\ddot{y} = \frac{C_f \rho_g U_\infty^2}{C_b \rho_l R_0^2} - \frac{C_k \sigma}{\rho_l R_0^3}y - \frac{C_d \mu_l}{\rho_l R_0^3}\dot{y},
\label{eq:tab_ode}
\end{align}
where $y=(R-R_s)/(C_b R_s)$ and $R_s=(3V_l/4\pi)^{1/3}$. The model coefficients are $C_f=1/3$ , $C_k=8$, $C_d=5$, and $C_b=0.5$.

To summarize, the steps for the V-TAB model are as follows: 1) Eq. \eqref{eq:tab_ode} is solved to obtain the temporal evolution of  $R$. 2) $A_w$ is estimated using Eq. \eqref{eq:Af_approx}. 3) The drop vaporization rate is then calculated using Eq. \eqref{eq:volume_evap_rate_modified}). The algorithm for the V-TAB model is also given in the Algorithm table \ref{tab:algor}.

\begin{algorithm}[H]
\caption{Vaporizing-TAB algorithm summary}\label{alg:cap}
\begin{algorithmic}
\State Initialize $R_0$
\State Calculate $\Delta t=\tau_c/10^3$
\While{$t<t_{f}$}
\State Compute $y$ using TAB Eq. \eqref{eq:tab_ode}
\State Determine $R=C_b R_s y+R_s$
\State Approximate $A_w$ using Eq. \eqref{eq:Af_approx} 
\State Compute the vaporization rate (Eq. \eqref{eq:volume_evap_rate_modified})
\State Update the drop volume, $V_l^{n+1}=V_l^{n}+\Delta t(dV_l/dt)$
\State Update the equivalent spherical radius, $R_s=(3V_l^{n+1}/4\pi)^{1/3}$
\EndWhile
\end{algorithmic}
\label{tab:algor}
\end{algorithm}

\begin{figure}[tbp]
	\begin{center}
		\includegraphics [width=1.0\textwidth]{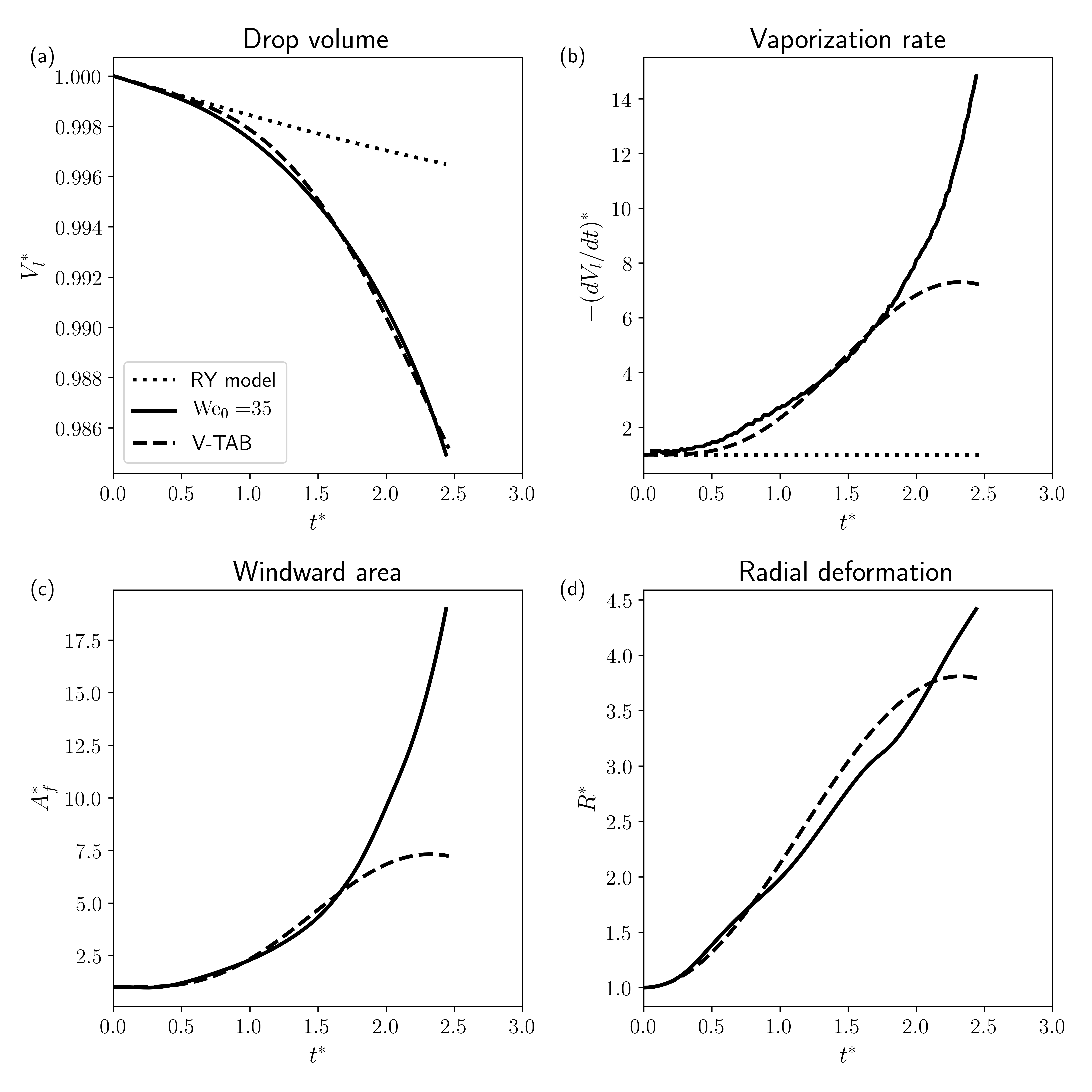}
	\end{center}
	\caption{Time evolutions of (a) the drop volume, (b) the vaporization rate, (c) the windward surface area, and (d) the relative velocity of the drop until breakup for $\text{We}_0=35$ at $L=14$. The V-TAB model is also shown for this case until $R^*$ reaches a maximum value.}
	\label{fig:V-TAB} 
\end{figure}

The V-TAB model predictions are compared to the simulation results for $\text{We}_0=35$ in Fig.~\ref{fig:V-TAB}. Until about $t^*=1.5$, the windward surface area $A_w^*$ and lateral radius $R^*$ predicted by the TAB model agree well with the simulation results (Fig. \ref{fig:V-TAB}(c)-(d)). Correspondingly, the V-TAB model well predicts the drop vaporization rate in this time period, see Fig. \ref{fig:V-TAB}(b). After $t^*=1.5$, the predicted $A_w^*$ deviates from the simulation results because the TAB model failed to capture the rapid increase of $R^*$ due to bag inflation. Additionally, the TAB model typically assumes breakup occurs when the drop deformation reaches $R^*=2$, which is an under-prediction in the drop breakup time. 

The results for the RY model for spherical drops are also plotted in Fig.~\ref{fig:V-TAB}(a) for comparison. Over the duration of the simulation $t^*\le 2.5$, the predicted evolution of the drop volume $V^*_{l}$ generally agrees much better with the simulation results than the RY model. The RY model significantly underestimates the drop vaporization rate. The V-TAB model predicts that the drop volume reduces by 0.015 of the original volume at $t^*\le 2.5$, while the volume loss measured in the simulation is 0.016. In contrast, the RY model predicts a volume loss of only 0.004. Therefore, the V-TAB model is far more accurate than the RY model in predicting the drop volume.

\begin{figure}[tbp]
	\begin{center}
		\includegraphics [width=1.0\textwidth]{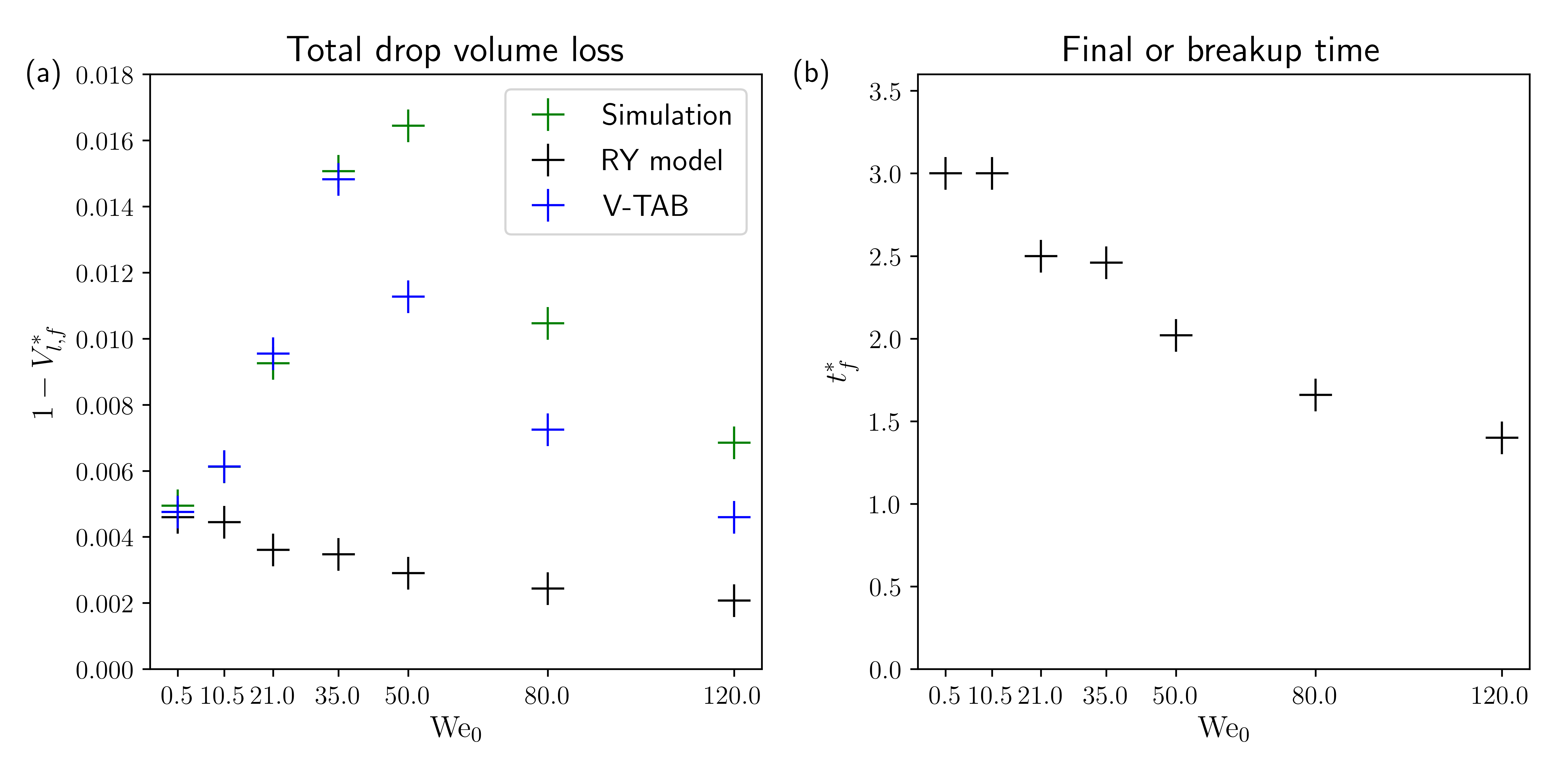}
	\end{center}
	\caption{(a) Model predictions for the final loss in drop volume ($1-V^*_{l,f}$) for different $\text{We}_0$. (b) The final time ($t^*_f$), at which the drop volumes are measured, is defined as the drop breakup time for $\text{We}_0>10.5$ and as $t^*_f=3$ for drops that do not break ($\text{We}_0\ge10.5$).}
	\label{fig:V-TAB_vs_sim} 
\end{figure}

The total reduction in drop volume for different $\text{We}_0$ and $\text{Re}_0$ is depicted in Fig.~\ref{fig:V-TAB_vs_sim}(a). The final time ($t_f^*$) is defined as the breakup time if a drop breaks and is taken to be $t^*=3$ if a drop does not break. The values of $t_f^*$ for different cases are shown in Fig.~\ref{fig:V-TAB_vs_sim}(b). It can be observed from Fig.~\ref{fig:V-TAB_vs_sim}(a) that both the RY and V-TAB models accurately predict the drop vaporization for very low Weber numbers (\ie, $\text{We}_0=0.5$) due to the negligible drop deformation. As $\text{We}_0$ increases, the V-TAB model predictions match the simulation results much better than the RY model for spherical drops. Some discrepancy between the V-TAB model and the simulation results is observed for very large $\text{We}_0$ (\ie, $\text{We}_0=50-120$), which is likely due to the increasing complexity in drop morphology that cannot be resolved by the TAB deformation model. Nevertheless, the V-TAB does a far better job of predicting drop vaporization than the RY model. For $\text{We}_0<50$ the volume reduction upon breakup increases with  $\text{We}_0$ due to the larger deformation and the resulting enhancement of vaporization. While $\text{We}_0$ continues to increase for $\text{We}_0>50$, the drop breakup time decreases rapidly and, as a result, the volume reduction due to vaporization decreases with  $\text{We}_0$.

\section{Conclusions}
\label{conclusions}

In the present study, the vaporization of a freely moving drop in a gas stream of its own vapor is investigated through direct numerical simulation. As the drop vaporization is driven by heat transfer between the drop and the gas stream, the drop vaporization rate is characterized by the Nusselt number.  Validation and verification studies are first performed for a drop with a very low Weber number, and the drop remains approximately spherical. The Nusselt number measured from simulations agrees well with the empirical correlation for two fluids (acetone and water) over a wide range of Reynolds and Stefan numbers. As the Weber number increases but remains below the critical Weber number, the drop experiences shape vibration in time, causing oscillations in the drop windward surface area and vaporization rate. The normalized drop vaporization rate increases with the windward surface area, as the temperature gradient is high and the local vaporization rate on the windward surface is generally high.

For moderate Weber numbers, where the drop breaks in the bag and multi-mode regimes, the morphological evolution of the drop revealed in the present simulations are consistent with former studies, \ie, the drop deforms into a backward bag, followed by a forward rim bag. The high-temperature vapor is entrained into the wake, resulting in an increase in the local vaporization rate near the leeward pole. The increase in the drop surface area generally enhances the drop vaporization, and the vaporization rate generally increases with an increase in the Weber number.

Additionally, we considered the influence of the Reynolds number on the vaporization enhancement due to drop deformation. It was found that the enhancement in vaporization due to the increase in surface area was reduced as the Reynolds number decreased. The reduction in the enhancement of vaporization with drop deformation was attributed to the change in wake structure and the resulting reduced vaporization on the leeward side: at a lower Reynolds number (\ie, $\text{Re}_0=50$) less high-temperature vapor was convected into the wake, which reduces the rate of vaporization.

Though the parametric simulations were performed using 2D axisymmetric simulations, fully 3D direct numerical simulations for representative cases are also performed to verify the 2D axisymmetric approximation. The 3D simulation results, such as the correlation between the vaporization rate and the drop surface area, showed good agreement with the 2D axisymmetric simulations for the range of parameters considered.

Based on the correlation between vaporization and drop deformation extracted from the present simulations, a transient drop vaporization model, \ie, the Vaporizing-TAB (V-TAB) model, is developed. While the drop deformation is represented by the TAB model, the effect of drop deformation on the vaporization rate is incorporated by an approximate correlation between the normalized drop vaporization rate and the windward surface area. The predictions made by the new V-TAB model and the conventional RY model are compared with the simulation results for different Weber numbers. While the RY model significantly underestimates the vaporization rate for drops with finite Weber numbers, the V-TAB model shows significantly better agreement with the simulation results.

\appendix{}
\section{Grid-refinement study}
\label{app:drop_converge}
\begin{figure}[tbp]
	\begin{center}
		\includegraphics [width=1.\columnwidth]{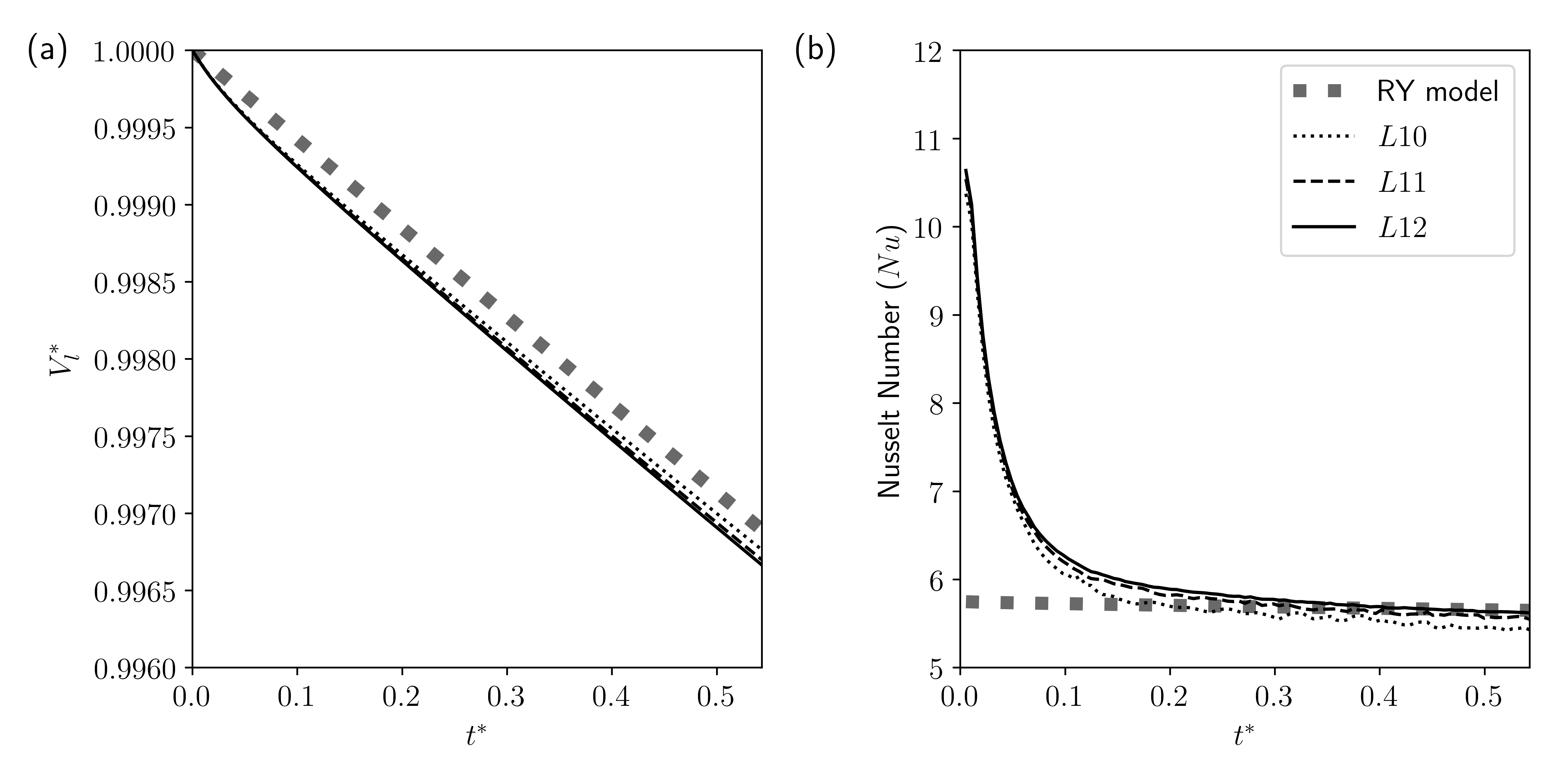}
	\end{center}
	\caption{Simulation results for (a) the normalized drop volume and (b) the Nusselt number for different mesh refinement levels. The results are for the acetone drop at $\text{Re}_0=60$, $\text{We}_0=0.5$, and $\text{St}=0.1$. The results of the RY empirical correlation (Eq.~\eqref{eq:volume_evap_rate_APPEN3}) are shown for comparison. }
	\label{fig:moving_drop_result} 
\end{figure}

The simulation results for the temporal evolution of the drop volume and Nusselt number are shown in Fig.~\ref{fig:moving_drop_result},
compared with the empirical model (Eq.~\eqref{eq:volume_evap_rate_APPEN3}). The simulation results converge as the refinement level increases from $L=10$ to $12$. The simulation results for $L=12$ and the empirical models agree very well after a quasi-steady state was achieved at $t^*>0.4$ or $t>5D_0/U_{\infty}$. The Nusselt number measured from the simulations is compared with the predictions of the empirical model in Table~\ref{tab:nu_drop} and an excellent agreement is observed.

 \begin{table*}[tb]
 \centering
\begin{tabular}{ll}
\hline
Mesh      & Nu   \\
 L=10       & 5.46 \\
 L=11       & 5.59  \\
 L=12       & 5.64 \\
 RY model & 5.65 \\
\hline
\end{tabular}

 \caption{The Nusselt numbers for various levels of grid refinement ($L=10$, $11$, $12$) compared to the Nu predicted by the RY model.}
 \label{tab:nu_drop}
 \end{table*}

\section{Sensitivity analysis for the initial temperature in the gas phase}
\label{app:R_inf}
\begin{figure}[tbp]
	\begin{center}
		\includegraphics [width=1.\columnwidth]{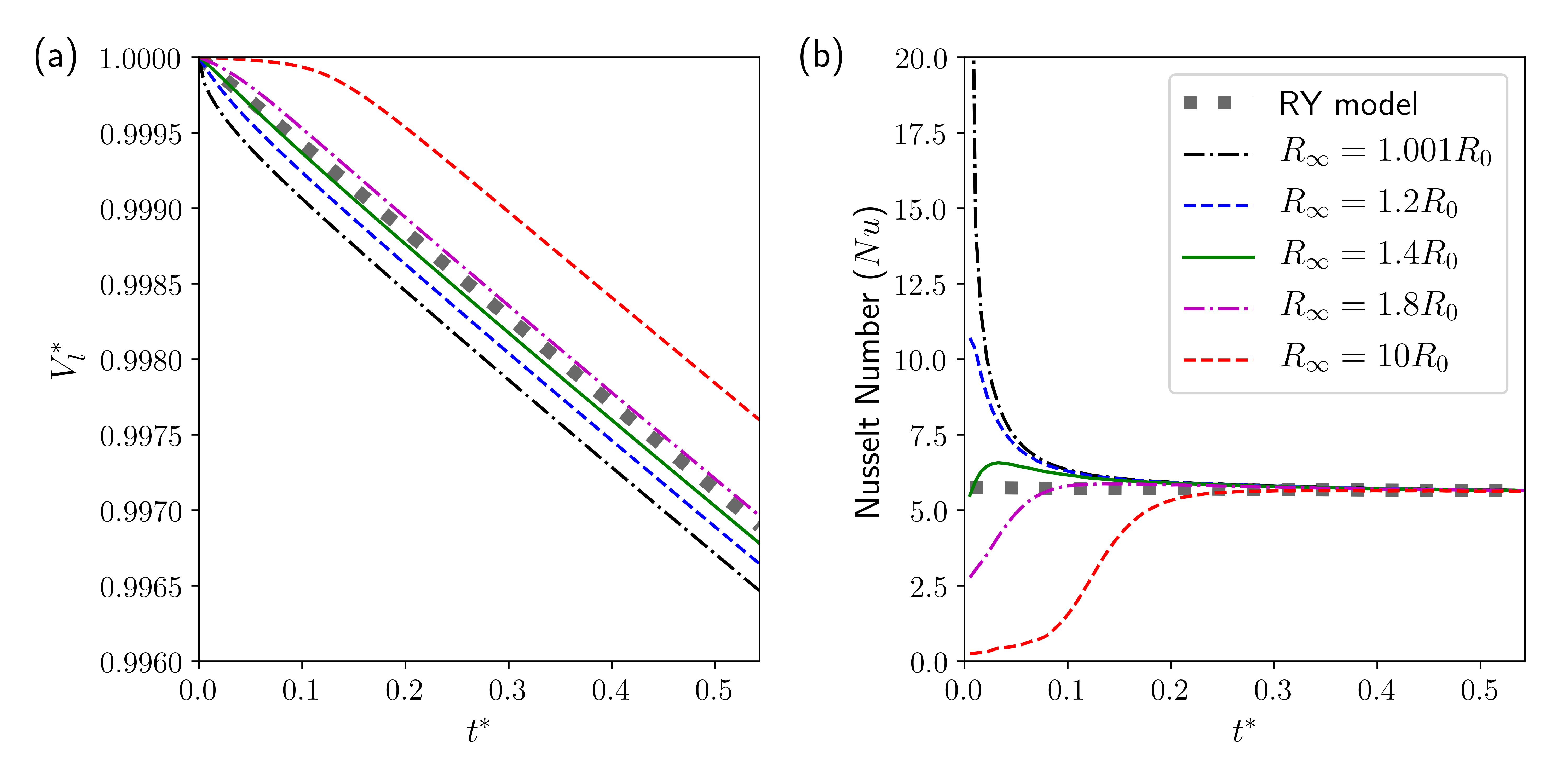}
	\end{center}
	\caption{Simulation results for (a) the normalized drop volume and (b) the Nusselt number for different initial temperature distributions surrounding the drop. The initial temperature distribution is modified using $R_{\infty}/$, where $R_{\infty}/R0=1.001$, $1.2$, $1.4$, $2$, $10$. The results are for the acetone drop at $\text{Re}_0=60$, $\text{We}_0=0.5$, and $\text{St}=0.1$. The results for the RY empirical correlation (Eq.~\eqref{eq:volume_evap_rate_APPEN3}) are shown for comparison.  }
	\label{fig:R_inf_result} 
\end{figure}

 To ensure that the quasi-steady state is independent of the initial temperature distribution, we varied $R_{\infty}$ which describes the initial condition (Eq. \eqref{eq:init_temp_drop}). Initially, the drop volume changes at different rates due to the different initial conditions, Fig. \ref{fig:R_inf_result}(a). However, as the flow develops, the Nusselt number approaches the same value in all cases, Fig. \ref{fig:R_inf_result}(b). At about $t^* \approx 0.4$ or $t_d^* \approx 5$, all cases appear to have reached the same quasi-steady state condition which is independent of the initial condition.

\section{Vaporization of a spherical drop driven by  heat transfer}
\label{app:V_dot}
The rate of drop volume loss due to vaporization depends on the heat flux to the surface of the drop, $q$
 \begin{align}
	\dot{V}_{l,s} = - \frac{q A_l}{\rho_l h_{l,g}}\,,
	\label{eq:volume_evap_rate_APPEN1}
\end{align}
where $h_{l,g}$ is the latent heat and the surface area of the spherical drop is $A_l=\pi D^2$, where $D$ is the drop diameter. The Nusselt number is defined as $\text{Nu}=D h/k_g$, where $h$ is the convective heat transfer coefficient. The heat flux to the drop can be expressed as a function of $h$, $q=h(T_{\infty}-T_\text{sat})$. 

With the dimensionless time ($t^*$) and drop volume ($V_l^*$) defined in Eqs.~\eqref{eq:t_star} and \eqref{eq:V_star}, the dimensionless vaporization rate can be written as  
 \begin{align}
	\dot{V}_{l,s}^* = \frac{d\, V_{l,s}^*}{d\, t^*} = - \frac{ 6 \text{Nu} \,\text{St}}{ \sqrt{\eta} \text{Re} \text{Pr}}\,. 
	\label{eq:volume_evap_rate_APPEN2}
\end{align}
If the RY empirical correlation, Eq. \eqref{eq:empirical}, is used for $\text{Nu}$, then Eq.~\eqref{eq:volume_evap_rate_APPEN2} becomes 
\begin{align}
	\dot{V}_{l,s}^* =  - \frac{6  [0.57 \text{Re}^{{1}/{2}}\text{Pr}^{{1}/{3}} + 2]}{ \sqrt{\eta} \text{Re} \text{Pr}}\frac{\text{St}}{(1 + \text{St})^{0.7}}\,.
	\label{eq:volume_evap_rate_APPEN3}
\end{align}

\section*{CRedit}
\textbf{Bradley Boyd}: Formal analysis, Methodology, Software, Validation, Visualization, Writing – original draft, Writing – review \& editing. \textbf{Sid Becker}: Writing – review \& editing, Supervision, Funding acquisition. \textbf{Yue Ling}: Methodology, Writing – review \& editing, Supervision, Funding acquisition.
\section*{Code availability}
All the codes necessary to simulate the vaporizing drops and boiling problems are available in \emph{Basilisk}’s sandbox \cite{boyd_code_2023}
\section*{Acknowledgments}
This research was supported by the ACS Petroleum Research Fund (\#62481-ND9) and the NSF (\#1942324). The authors also acknowledge the Extreme Science and Engineering Discovery Environment (XSEDE) and the Texas Advanced Computing Center (TACC) for providing the computational resources that have contributed to the research results reported in this paper. The Baylor High Performance and Research Computing Services (HPRCS) have been used to process the simulation results. The newly developed methods have been implemented in the open-source multiphase flow solver \emph{Basilisk}, which is made available by St\'ephane Popinet and other collaborators.


\bibliography{references}

\end{document}